\documentclass{aastex631}
\shorttitle{Hot subdwarfs with composite spectrum}
\shortauthors{Lei et al.}
\graphicspath{{./}{figures/}}

\begin{document}

\title{Hot subdwarf stars identified in LAMOST DR8 with single-lined and composite spectra}

\author[0000-0003-2362-6607]{Zhenxin Lei}
\affiliation{Key Laboratory of Stars and Interstellar Medium, 
Xiangtan University, Xiangtan 411105, People's Republic of China}
\affiliation{Physics Department, Xiangtan University, Xiangtan, 411105, People’s Republic of China}
\email{leizhenxin2060@163.com}

\author{Ruijie He}
\affiliation{Key Laboratory of Stars and Interstellar Medium, 
Xiangtan University, Xiangtan 411105, People's Republic of China}
\affiliation{Physics Department, Xiangtan University, Xiangtan, 411105, People’s Republic of China}

\author[0000-0003-0963-0239]{P\'eter N\'emeth}
\affiliation{Astronomical Institute of the Czech Academy of Sciences, CZ-251\,65, Ond\v{r}ejov, Czech Republic}
\affiliation{Astroserver.org, F\H{o} t\'er 1, 8533 Malomsok, Hungary} 

\author[0000-0001-6172-1272]{Joris Vos}
\affiliation{Astronomical Institute of the Czech Academy of Sciences, CZ-251\,65, Ond\v{r}ejov, Czech Republic}

\author{Xuan Zou}
\affiliation{Key Laboratory of Stars and Interstellar Medium, 
Xiangtan University, Xiangtan 411105, People's Republic of China}
\affiliation{Physics Department, Xiangtan University, Xiangtan, 411105, People’s Republic of China}

\author{Ke Hu}
\affiliation{Key Laboratory of Stars and Interstellar Medium, 
Xiangtan University, Xiangtan 411105, People's Republic of China}
\affiliation{Physics Department, Xiangtan University, Xiangtan, 411105, People’s Republic of China}

\author{Huaping Xiao}
\affiliation{Key Laboratory of Stars and Interstellar Medium, 
Xiangtan University, Xiangtan 411105, People's Republic of China}
\affiliation{Physics Department, Xiangtan University, Xiangtan, 411105, People’s Republic of China} \email{hpxiao@xtu.edu.cn}

\author{Huahui Yan}
\affiliation{Shandong Provincial Key Laboratory of Optical Astronomy and Solar-Terrestrial Environment, School of Space Science and Physics, Shandong University, Weihai 264209, China}

\author[0000-0003-2868-8276]{Jingkun Zhao}
\affiliation{Key Laboratory of Optical Astronomy, National Astronomical Observatories, Chinese Academy of Sciences, Beijing 100012, China}



\begin{abstract}
222 hot subdwarf stars were identified with LAMOST DR8 spectra, among which 131 stars show composite spectra and have been decomposed, while 91 stars present single-lined spectra. 
Atmospheric parameters of all sample stars were obtained by fitting Hydrogen (H) and Helium (He) line profiles with synthetic spectra. 
Two long-period composite sdB binaries were newly discovered by combining our sample with the non-single star data from \textit{Gaia} DR3. 
One of the new systems presents the highest eccentricity (i.e., $0.5\pm0.09$) among known wide sdB binaries, which is beyond model predictions. 
15 composite sdB stars fall in the high probability binary region of RUWE-AEN plane, and deserve priority follow-up observations to further study their binary nature. 
A distinct gap is clearly presented among temperatures of cool companions for our composite-spectra sample. But we could not come to a conclusion whether this feature is connected to the formation history of hot subdwarf stars before their binary natures are confirmed.
\end{abstract}



\section{Introduction} \label{sec:intro}
Hot subdwarf stars have similar spectra as O/B type main-sequence (MS) stars, but at lower luminosity. 
They have low masses around at 0.5 M$_{\odot}$ but very high effective temperatures and surface gravity (e.g., roughly 20000 K $\leq$ $T_\mathrm{eff}$  $\leq$ 70000 K  and  5.0 $\leq$ $\mathrm{log}\ g$   $\leq$ 6.5, see \citealt{2009ARA&A..47..211H,2016PASP..128h2001H} for excellent review on this type of stars). 
Hot subdwarf stars are burning helium (He) in their cores or have evolved even off this stage. 
Many of these stars occupy the bluest positions at the end of the horizontal branch (HB) in the Hertzsprung-Russell (H-R) diagram, hence they are also known as extreme horizontal branch (EHB) stars, which show up most remarkably in globular clusters (GCs). Some other types of stars could also go across the hot subdwarf region in the H-R diagram \citep{1986A&A...155...33H,2016PASP..128h2001H}, such as post-EHB stars, post asymptotic giant branch (post-AGB) stars, low and extremely low mass (pre-)white dwarfs (pre-ELM WDs), etc.

According to the characteristics of spectral lines, hot subdwarf stars can be classified into several sub types, e.g., sdB, sdOB, sdO, He-sdB, He-sdOB and He-sdO \citep{1990A&AS...86...53M, 2017A&A...600A..50G}. sdB stars have dominant H Balmer lines but weak or absent He lines. sdOB stars have  dominant H Balmer lines, together with both weak He I and He II lines, while sdO stars present dominant H Balmer lines and an obvious He II line at 4686 \AA, but without obvious He I lines.  On the other hand, He-sdB stars show dominant He I lines, but weak or absent H Balmers and He II lines. He-sdOB stars present both strong He I and He II lines, but with weak or absent H Balmer lines, while He-sdO stars present  strong He II lines, but weak or absent H Balmer lines and HeI lines.

Since many of hot subdwarfs were found in close binary systems, they are considered to be mainly formed in binary evolution. 
\citet{2002MNRAS.336..449H,2003MNRAS.341..669H} conducted detailed binary population synthesis and found that stable Roche lobe overflow (RLOF), common envelope (CE) ejection and a merger of two He white dwarf (WD) stars in binary evolution are three main possible formation channels for sdB stars (also see \citealt{2013MNRAS.434..186C}). \citet{2011ApJ...733L..42C} proposed that the merger of an He WD with an M dwarf star can produce single sdB stars. This model could account for the narrow mass range of single sdB stars and the existence of long-period sdB+MS binaries. On the other hand,  \citet{2012MNRAS.419..452Z} studied the merger of two He WDs in short orbital period, and found that both fast and slow accretion processes could reproduce He-rich hot subdwarf stars. Moreover, \citet{2017ApJ...835..242Z} also found that the merger of a He WD and a main-sequence star could contribute to the formation of intermediate He-rich subdwarfs. \citet{2015MNRAS.449.2741L,2016MNRAS.463.3449L} also found that tidally-enhanced stellar wind in binary evolution could contribute on the formation of blue hook (BHk) stars in GCs. 

Long period composite sdB binaries are the result of the stable-RLOF formation channel. In fact, F/G/K type companions are only found in long period sdB binaries. Their existence was predicted by \citet{2001ASPC..226..192G}, but it had taken another decade until the first orbits of long period sdBs were published \citep{2012MNRAS.421.2798D, 2012ASPC..452..163O, 2012A&A...548A...6V}. Currently 23 composite sdB binaries have solved orbits. These systems are found with orbital periods ranging from 2 to well over 4 years, and their properties have several puzzling features \citep{2019MNRAS.482.4592V}. Tidal circularisation theory predicts the sdB progenitors to circularize completely before the interaction phase starts, however, most observed wide sdB binaries have eccentric orbits. The interaction with a circumbinary disk during the mass loss phase can explain many, but not all, of these eccentric systems \citep{2015A&A...579A..49V}. More recently a strong correlation between orbital period and mass-ratio was linked to the galactic metallicity evolution \citep{2020A&A...641A.163V}. Several other peculiarities, as for example the bimodal distribution in the period - mass ratio and period - eccentricity parameter space, remain unexplained. These composite sdBs are a valuable population to study binary interaction theories as the orbital properties and the spectral properties of both components can be solved, and because all systems form a single population.

From an observational point of view, with the huge data released from large photometric and spectroscopic surveys, new discoveries of hot subdwarf stars have increased very quickly in recent years. 
A catalog of known hot subdwarf stars was compiled from literature and reported by \citet{2017A&A...600A..50G}, which contains 5613 objects and a lot of useful information, such as: multi-band photometry, ground based proper motions, classifications, published atmospheric parameters, etc. 
\citet{2019MNRAS.486.2169K} confirmed 20\,088 WDs, 425 hot subdwarfs, and 311 cataclysmic variables (CVs) based on the spectra released by Sloan Digital Sky Survey Data Release 14 (SDSS DR14, \citealt{2018ApJS..235...42A}). 
On the other hand, the second data release of the \textit{Gaia} mission (\textit{Gaia} DR2) became available to the public in 2018, which provided precise astrometry and photometry for more than 1.3 billion sources over the full sky \citep{2018A&A...616A...1G}. 
With this useful information, hot subdwarf candidates could be selected more efficiently and accurately in H-R diagram \citep{2018A&A...616A..10G}. 
Based on \textit{Gaia} DR2 data, \citet{2019A&A...621A..38G} compiled a candidates catalog by means of colour, absolute magnitude, and reduced proper motion cuts, in which 39\,800 hot sub-luminous candidates were selected, and it proved as a good input catalog for large photometric and spectroscopic surveys for further study of hot subdwarfs. The number of candidates increased to 61\,585 by \citet{2022A&A...662A..40C} as new data was updated from \textit{Gaia} EDR3 \citep{2021A&A...649A...1G}. 

With the help of precise photometric magnitudes and parallaxes provided by \textit{Gaia} DR2, more than 1000 hot subdwarfs were identified in The Large Sky Area Multi-Object Fiber Spectroscopic Telescope (LAMOST) spectral survey \citep{2019ApJ...881....7L,2021ApJS..256...28L,2018ApJ...868...70L,2019ApJ...881..135L,2020ApJ...889..117L}, and reliable  atmospheric parameters (e.g., effective temperature, gravity and He abundance) were obtained by fitting hydrogen (H) and He line profiles with synthetic spectra. 
Moreover, \citet{2020A&A...635A.193G} compiled a new catalog with 5874 known hot subdwarf stars recorded, including 528 newly discovered objects and removing 268 previously misclassified objects with respect to the previous catalog of \citet{2017A&A...600A..50G}. 

On the other hand, artificial intelligence (AI) methods were widely used recently in research fields of astronomy and astrophysics, including searching for hot subdwarfs in large photometric and spectroscopic surveys. 
\citet{2017ApJS..233....2B} firstly selected about 7000 hot subdwarf candidates from the first data release (DR1) of LAMOST spectral survey, by employing a machine learning algorithm. 
Based on these candidates, \citet{2019PASJ...71...41L} identified 56 hot subdwarf stars by detailed spectral analysis, and provided reliable atmospheric parameters. 
Subsequently, a deep learning method equipped by convolutional neural networks and a support vector machine (CNN+SVM) was used in \citet{2019ApJ...886..128B} to improve the accuracy and efficiency of hot subdwarf candidate selection in LAMOST spectral dataset. 
Furthermore, \citet{2022ApJS..259....5T} selected 2393 hot subdwarf candidates from LAMOST DR7-V1 dataset by using a robust identification method based on CNN, and confirmed 2067 hot subdwarf stars, including 25 new discoveries. 

Though a large number of hot subdwarf stars were discovered profiting from the huge data releases of large photometric and spectroscopic surveys, most of them were identified by single-lined spectra, while hot subdwarf systems showing composite spectra were rarely reported because their analysis requires complicated decomposition processes.
\citet{2012MNRAS.427.2180N} decomposed 29 hot subdwarf composite spectra from Galaxy Evolution Explorer (GALEX) survey, and investigated the incidence of A, F and G type companions. 
\citet{2018MNRAS.473..693V} presented 148 composite sdB systems retrieved from literature, but only 9 of them were decomposed from spectra obtained by  Ultraviolet and Visual Echelle Spectrograph (UVES) at VLT. Furthermore, 
\citet{2021A&A...653A...3N} decomposed a long-period sdOB+G1V type composite binary SB 744. They found that sdOB primary of the binary presented over abundances of lead and fluorine, and confirmed it as an old, Population II and heavy-metal system in the Halo.   
\citet{2021A&A...653A.120D} also decomposed a long-period binary, which contains a lead-rich sdOB star and a metal-poor subdwarf F-type (sdF) companion, and obtained their atmospheric parameters by performing a detailed analysis of high-resolution SALT/HRS and VLT/UVES spectra.

In previous studies \citep{2018ApJ...868...70L, 2019ApJ...881..135L, 2020ApJ...889..117L}, we identified more than 800 hot subdwarf stars with LAMOST spectra, but composite-spectrum candidates were not analyzed.
In this study, we selected 257 hot subdwarf candidates with composite spectra from the  LAMOST DR8 dataset, and identified 131 hot subdwarf stars through  decomposition processes. 
Moreover, 91 hot subdwarf stars were confirmed from 203 single-lined spectra. 
The structure of this study is as follows: In Section 2, we described how hot subdwarf candidates were selected from LAMOST DR8 with the help of \textit{Gaia} EDR3 H-R diagram; 
The method for decomposition of composite spectra is illustrated in Section 3; 
Our results are given in Section 4, while a discussion together with a summary closed this paper.

\section{Sample selection}

\begin{figure}

    \centering
    \begin{minipage}[c]{0.49\textwidth}
    \includegraphics [width=90mm]{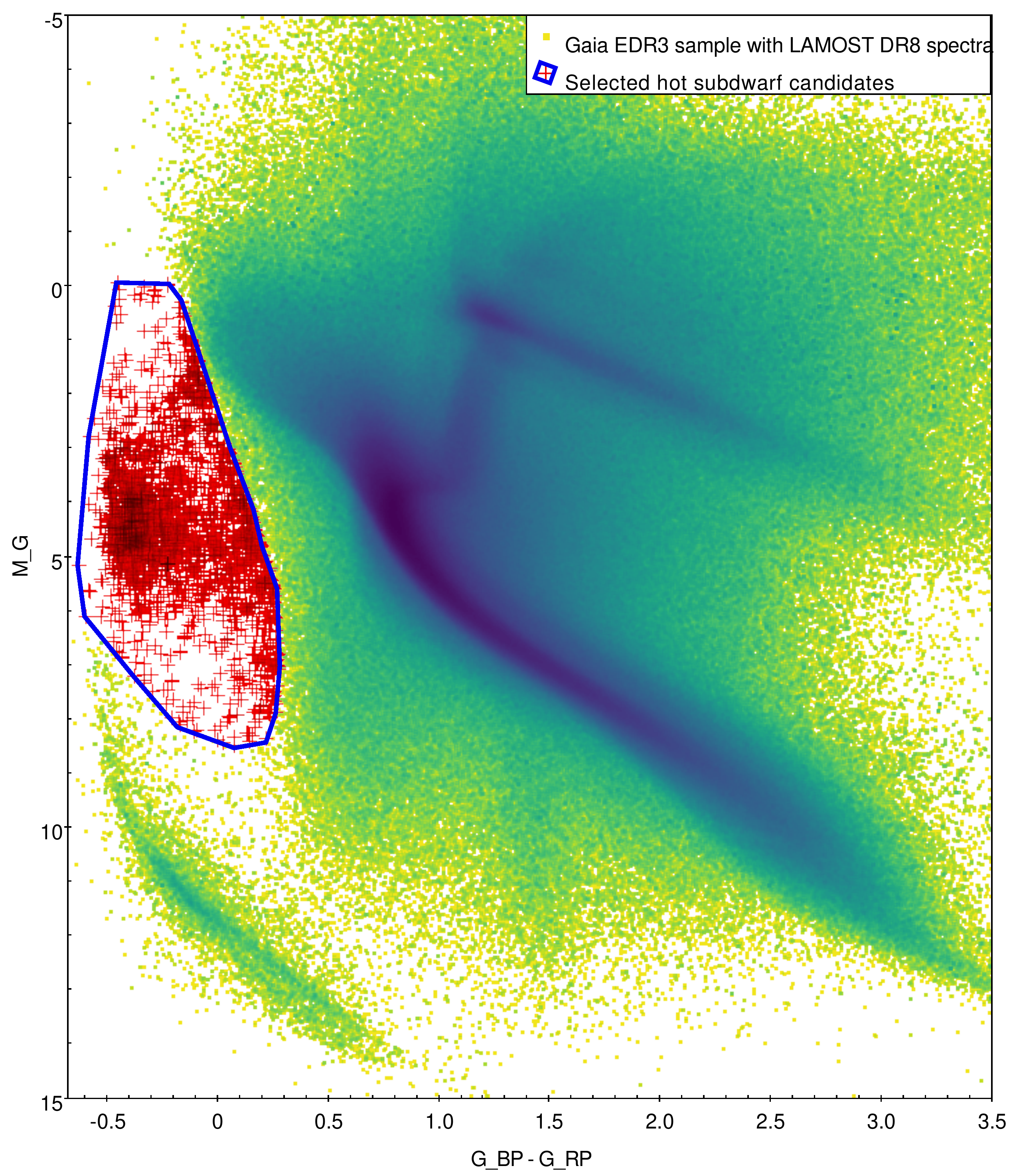}
    \end{minipage}%
    \begin{minipage}[c]{0.49\textwidth}
    \includegraphics [width=90mm]{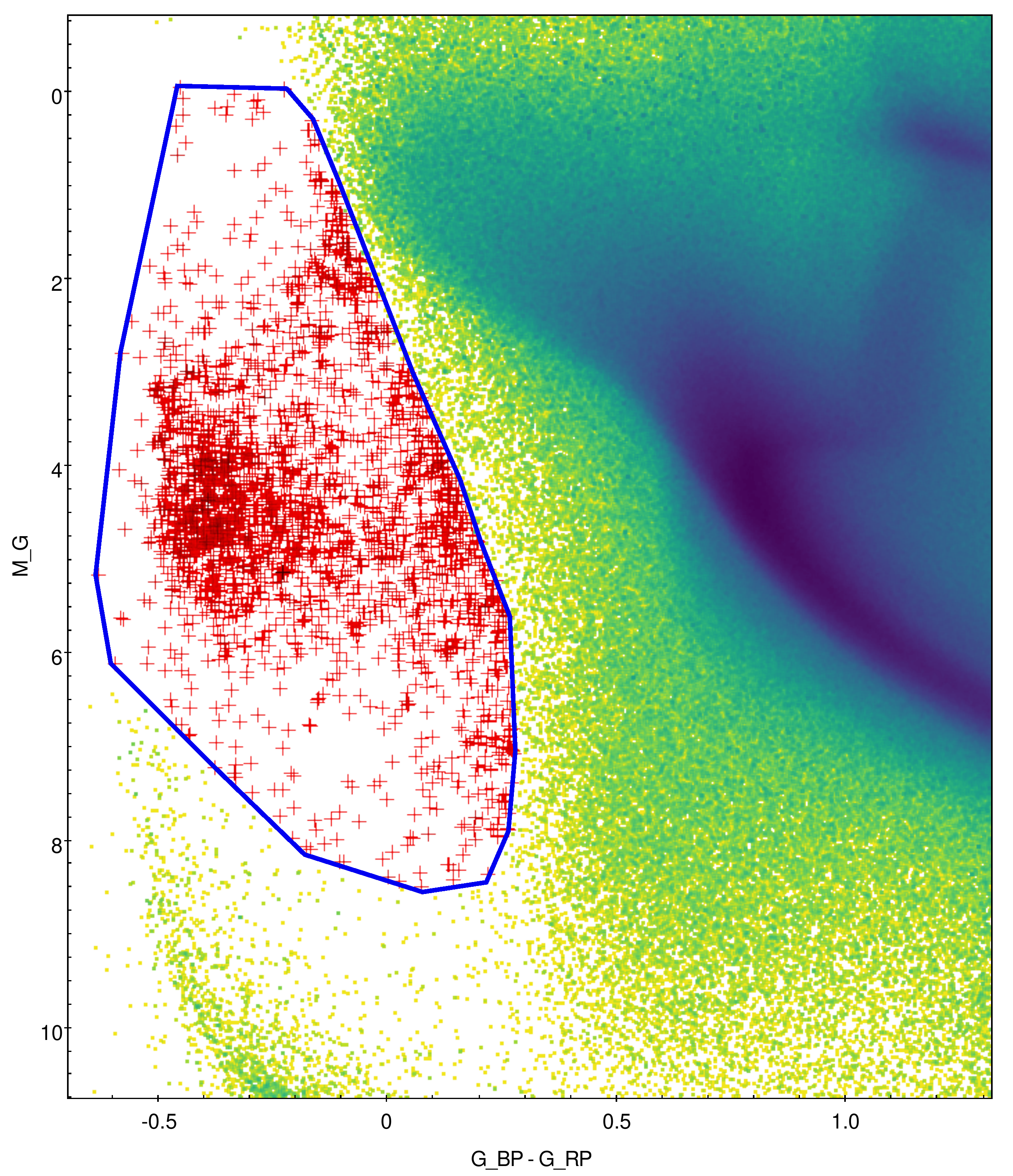} 
    \end{minipage}
    \caption{Left panel: Hot subdwarf candidates selection in \textit{Gaia} EDR3 H-R diagram. \textit{Gaia} EDR3 sample with LAMOST DR8 low-resolution spectra are represented by filled squares, while 3740 candidates selected through a  manually drawn polygon (blue solid line) around the hot subdwarf area are denoted by red  pluses. Right panel: Zoom in the hot subdwarf area.}
    \label{fig1}
\end{figure}

\subsection{\textit{Gaia} EDR3 and \textit{LAMOST} DR8}
The early installment of the third Gaia data release (\textit{Gaia} EDR3, \citealt{2021A&A...649A...1G}) was opened to the public on December 2020, which contains the first 34 months of data collection of the \textit{Gaia} mission \citep{2016A&A...595A...1G}. 
It consists of 1.8 billion objects brighter than 21 magnitude with precise astrometry, as well as photometry in the $G$, $G_{BP}$, and $G_{RP}$ bands.  
It also presents an updated list of radial velocity data (RV) from \textit{Gaia} DR2 \citep{2018A&A...616A...1G} after removing some spurious values. 
\textit{Gaia} EDR3 made significant improvements in precision and accuracy of photometric magnitudes, parallax, celestial positions and proper motions,  with respect to \textit{Gaia} DR2. With these useful information, H-R diagram became available based on a huge number of stellar objects (e.g., see \citealt{2018A&A...616A..10G}), and it significantly improved the efficiency of candidate selection for hot subdwarfs (see \citealt{2018ApJ...868...70L} for a detailed discussion).

The Large Sky Area Multi-Object Fiber Spectroscopic Telescope (LAMOST) is a special reflecting Schmidt telescope with 4000 fibers in a field of view of 20 $deg^2$ in the sky \citep{2012RAA....12.1197C, 2012RAA....12..723Z, 2006ChJAA...6..265Z}, which is operated by the National Astronomical Observatories of the Chinese Academy of Sciences. LAMOST launched its pilot survey in October 2011, and completed the eighth regular survey in May 2020. All the data collected during this period consists of the eighth data release (DR8) of LAMOST. LAMOST DR8 released a total number of 11,214,076 low resolution spectra, including 10,388,423 stellar spectra, 219,776 galaxy spectra, 71,786 quasar spectra, and 534,091 spectra of unknown objects. These spectra cover the wavelength range of 3690 - 9100\AA\  with a resolution of 1800 at 5500 \AA.

Since October 2018, LAMOST also started its stage II survey, which contains both low- (e.g., R=1800 at 5500 \AA) and medium-resolution (e.g., R=7500 at 5163 \AA\ and 6593 \AA) spectroscopic surveys. For the medium-resolution survey, 1,479,145 coadded spectra and 4,559,091 single exposure spectra were collected by non-time-domain and time-domain surveys, respectively. Both low- and medium-resolution spectra are available on the LAMOST DR8 website \footnote{\url{http://www.lamost.org/dr8/v1.0/}}.

\subsection{Candidates selection}
To select hot subdwarf candidates from LAMOST DR8, we cross-matched the LAMOST DR8 low-resolution dataset with the \textit{Gaia} EDR3 dataset, and obtained 10,915,264 common sources. 
In the left panel of Fig \ref{fig1}, all common sources are plotted in the \textit{Gaia} H-R diagram with filled squares, while 
our candidates were selected from a manually drawn polygon around the hot subdwarf region (blue solid line). Red pluses in the polygon denote 3740 hot subdwarf candidates selected from \textit{Gaia} EDR3, which also have LAMOST low-resolution spectra. The right panel zooms in the hot subdwarf region for clarity.

\begin{figure}
    \centering
    \includegraphics[width=110mm]{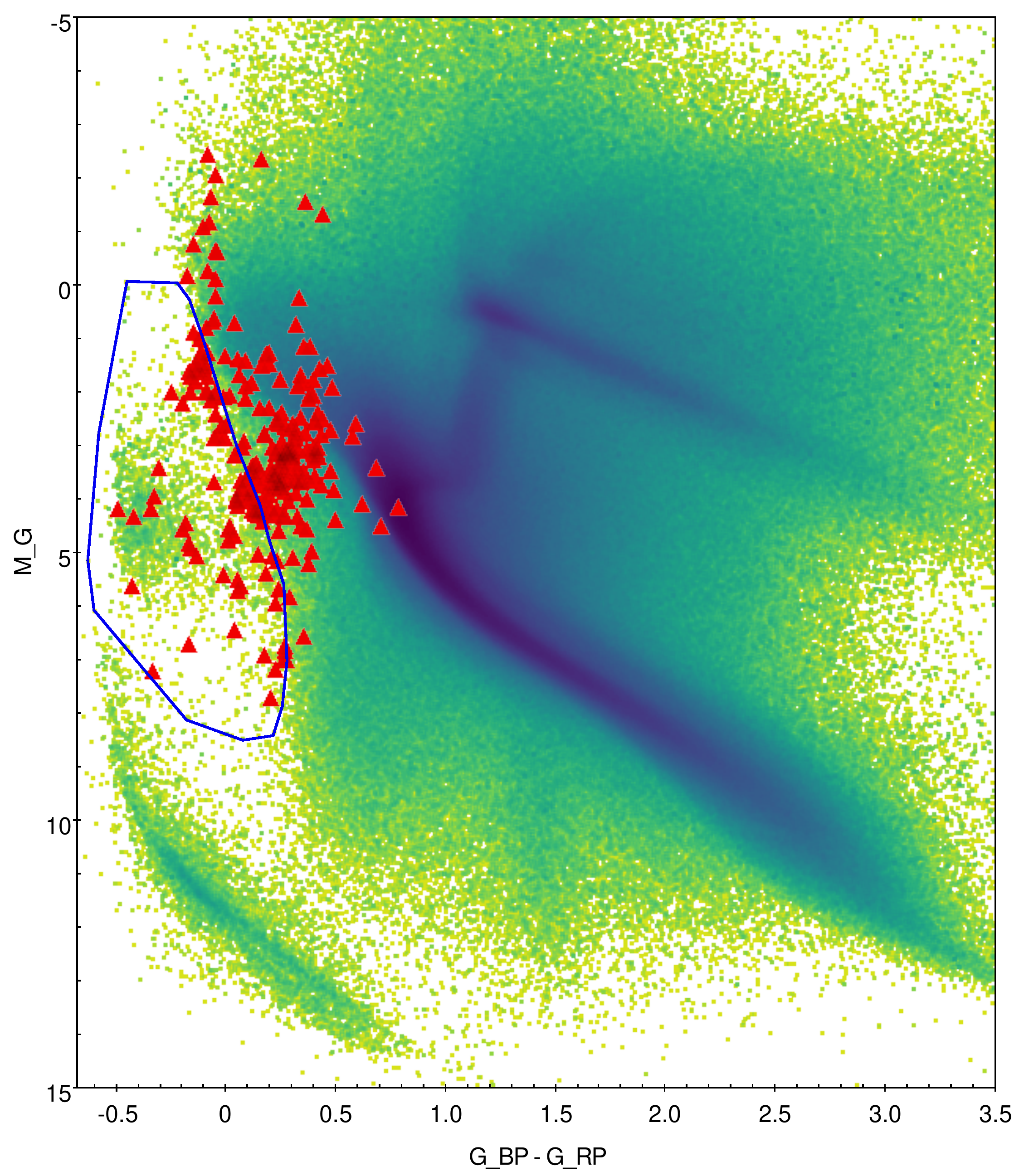}
    \caption{Composite-spectra hot subdwarf candidates in \textit{Gaia} EDR3 H-R diagram. \textit{Gaia} EDR3 sample with LAMOST DR8 low-resolution spectra are represented by filled squares, while the composite-spectra candidates selected both from \textit{Gaia} EDR3 and the catalog of \citet{2019A&A...621A..38G} are shown by red filled triangles. The blue solid line presents the candidates-selection polygon in Fig \ref{fig1}.}
    \label{fig2}
\end{figure}

In order to include as many candidates as possible, we also cross-matched our LAMOST DR8 dataset with the catalog of \citet{2019A&A...621A..38G}, which consists of 39\,800 hot subdwarf candidates selected from \textit{Gaia} DR2 (see Section 1),  and obtained 2700 common objects. 
We combined the two parts of candidates mentioned above together and obtained 6440 candidates in total. 
After removing duplicate sources, sources with signal to noise ratio in the \textit{u} band (SNRU) less than 10.0, and common candidates analyzed in our previous studies \citep{2018ApJ...868...70L,2019PASJ...71...41L,2019ApJ...881..135L,2020ApJ...889..117L}, 460 hot subdwarf candidates remained for further spectral analysis, among which 257 have composite spectra (e.g., having obvious Mg I triplet lines at 5170 \AA \ and/or Ca II triplet lines at 8650 \AA), while 203 have single-lined spectra. 
Different from our previous work in which only single-lined spectra were analyzed, atmospheric parameters were derived by spectral analysis both for composite and single-lined spectra in this study. Fig \ref{fig2} presents the 257 candidates with composite spectra by red filled triangles in \textit{Gaia} EDR3 H-R diagram. Because the candidates in the catalog of \citet{2019A&A...621A..38G} were selected by three different criteria, i.e., color cut, absolute magnitude and reduced proper motion (see Section 3 in their study for details), and because interstellar extinction was not considered for their \textit{Gaia} EDR3 sample,  many of the composite-spectra candidates in Fig \ref{fig2} are beyond the blue polygon of Fig \ref{fig1}, or even located in the MS area.

\section{spectral analysis}

\begin{figure}
    \begin{minipage}[]{12cm}      
    \includegraphics[height =7cm, width = 16cm]{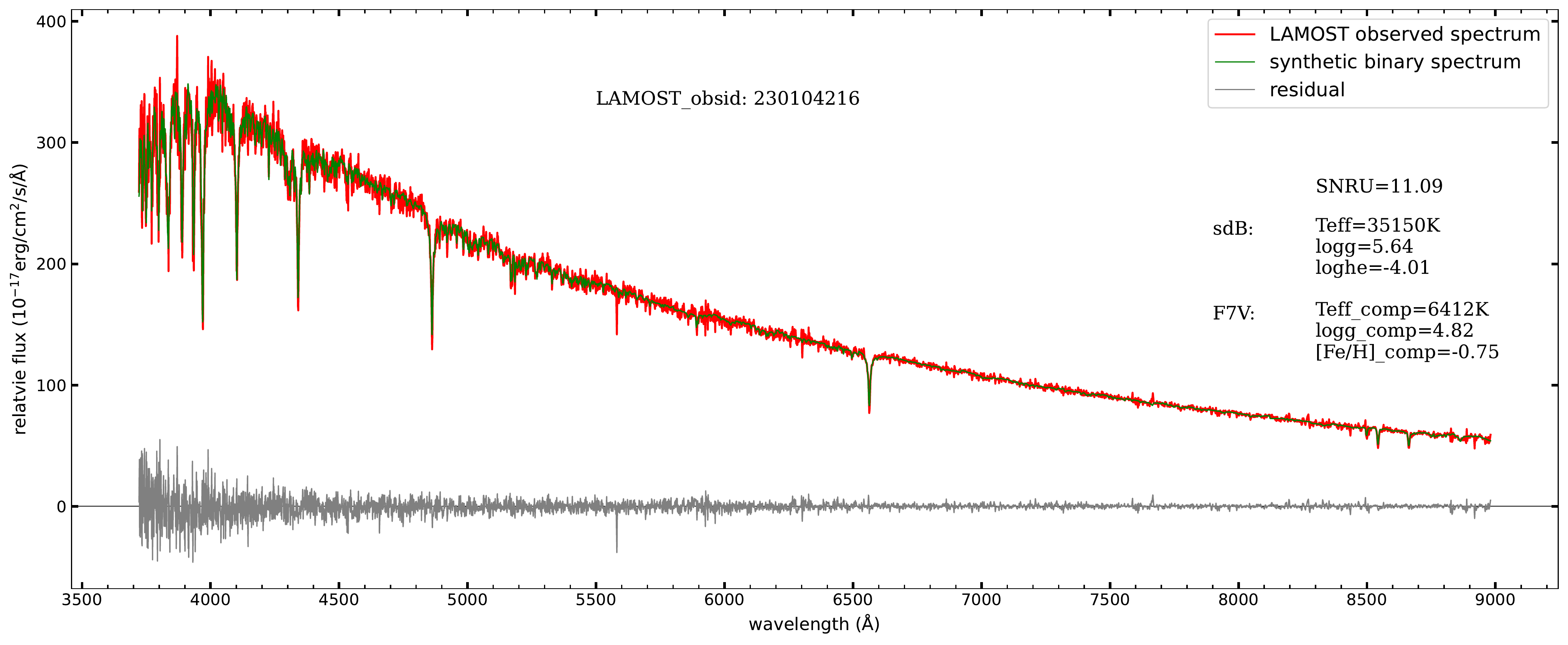} 
    \end{minipage}\\
    \begin{minipage}[]{12cm}      
    \includegraphics[height = 7cm, width = 16cm]{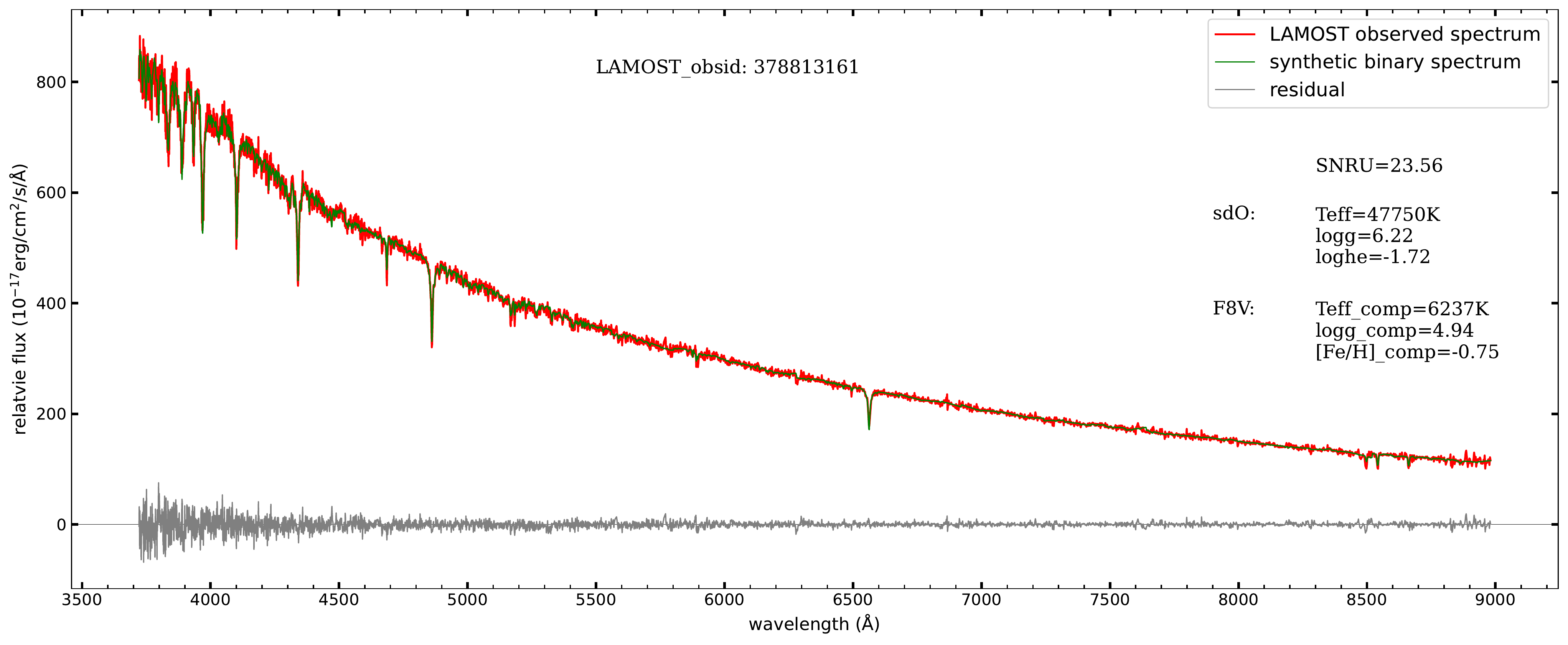} 
    \end{minipage}\\
    \begin{minipage}[]{12cm}      
    \includegraphics[height = 7cm, width = 16cm]{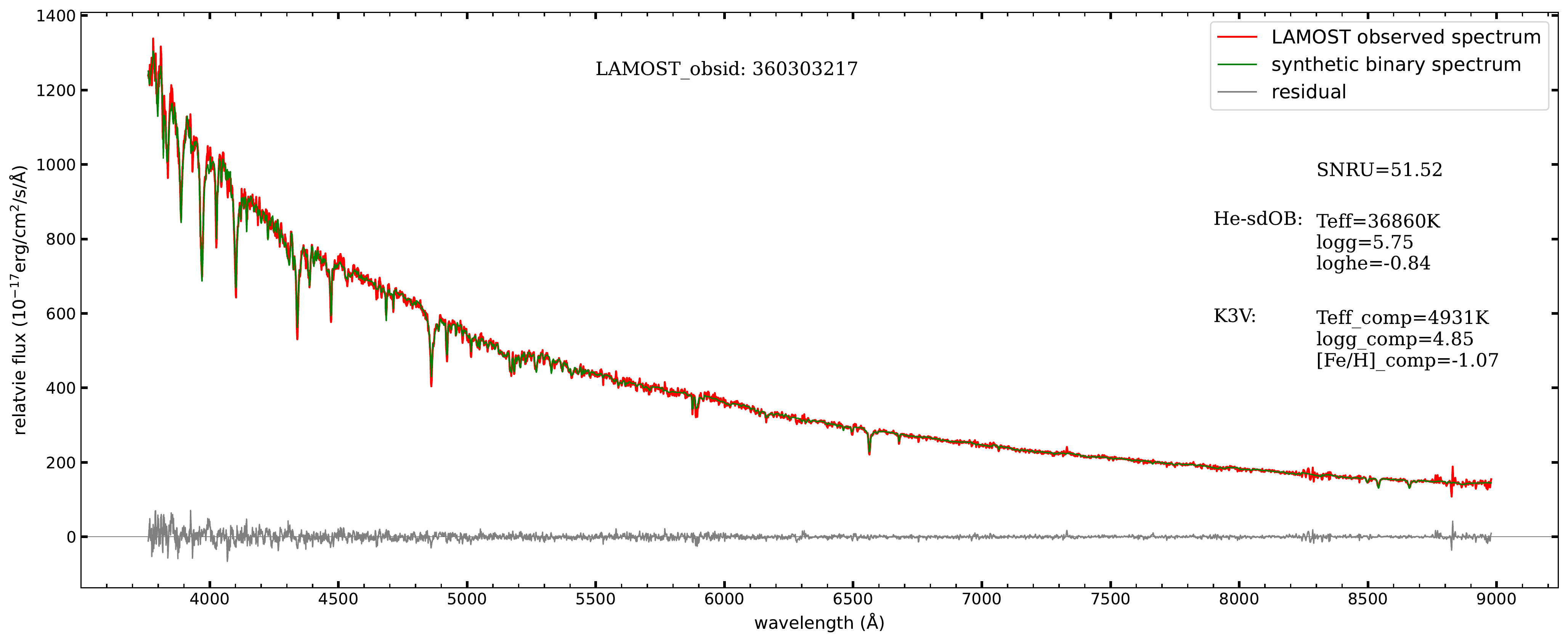} 
    \end{minipage}
    \caption{Fitting examples for three composite spectra. Observed spectra are presented by red curves, synthetic binary spectra combined a hot subdwarf spectrum and a cool stellar spectrum are denoted by green curves, while the residuals between observed and synthetic spectra are shown by gray curves. The signal-to-noise ratio in \textit{u} band (SNRU) of the observed spectra and the atmospheric parameters for the best-fitting synthetic binary spectra are also shown on the right side of each panel.}  
    \label{fig3}
\end{figure}

We used the spectral analysis tool, {\sc XTgrid} \citep{2012MNRAS.427.2180N,2014ASPC..481...95N}, to analyze both composite and single-lined spectra. 
The observed hot subdwarf spectra were reproduced by {\sc Synspec} synthetic spectra (version 49; \citealt{2007ApJS..169...83L}) calculated from non-LTE {\sc Tlusty} model atmospheres (version 204; \citealt{2017arXiv170601859H}).
We used the same steepest-descent $\chi^2$ minimization method implemented in {\sc XTgrid} to fit single-lined spectra in previous work, and recommend the readers to see  \citet{2018ApJ...868...70L,2019ApJ...881..135L,2020ApJ...889..117L} for detailed descriptions. 
The analysis of composite spectra is analogous, but we apply a linear combination of a synthetic hot subdwarf spectrum calculated from {\sc Tlusty} models and a cool companion spectrum interpolated from  ATLAS9 spectra \citep{2012AJ....144..120M, 2017AJ....153..234B} to render theoretical composite spectra.
Then, we iteratively modify the stellar parameters of both components and their flux contribution to the composite spectrum to fit an observation. 
Once the relative changes of all free parameters and the $\chi^2$ are below 0.5\%, we consider the fit converged. Next, error calculations are performed in one dimension, changing each parameter until the $\chi^2$ reaches the corresponding change for 60\% confidence. 

The spectral modeling of composite spectra is heavily affected by parameter correlations and large systematics, therefore photometric data were collected from the VizieR Photometry Viewer\footnote{\url{http://vizier.u-strasbg.fr/vizier/sed/}} service within 2 arcseconds around the targets and the spectral energy distribution (SED) was built from these data. The SED was used to optimize the flux contributions of the binary members to the composite continuum.

Fig \ref{fig3} gives fitting examples for three composite spectra. In the top panel, the observed spectrum (LAMOST\_obsid=230104216, red curve) was decomposed to an sdB type spectrum and an F7V type cool spectrum (green curve). 
The middle panel shows how an sdO type spectrum combined with an F8V type spectrum matched the observed spectrum of LAMOST\_obsid=378813161. 
The spectrum of LAMOST\_obsid=360303217 was decomposed to a He-sdOB type spectrum and a K3V type spectrum in the bottom panel. 
The parameters of the synthetic spectra for the hot and cool components are presented on the right side of each panel.  

\section{results}
By employing the method described in Section 3, 222 hot subdwarf stars were identified, among which 131 stars were identified with composite spectra (see Table 1), and 91 stars with single-lined spectra (see Table 2). 
We cross-matched our samples with the hot subdwarf catalog of \citet{2020A&A...635A.193G} and found 148 common objects. 
However, among the 148 common stars, there are 109 objects without atmospheric parameters in the catalog of \citet{2020A&A...635A.193G}. 
It means that our study not only presents 74 new discoveries, but also provides atmospheric parameters for 109 hot subdwarfs that were previously identified but not yet analyzed with atmosphere models. 

\subsection{Main property table for the samples}

\begin{longrotatetable}
\begin{deluxetable*}{@{\extracolsep{2pt}}lllclllllllllllll@{}}
\tablecaption{131 hot subdwarf stars identified with composite spectra.}
\setlength{\tabcolsep}{1mm}{} 
\tablewidth{100pt}
\tabletypesize{\tiny}
\tablehead{
&\multicolumn{4}{c}{Composite system}&&& &\multicolumn{4}{c}{Hot primary} &&\multicolumn{4}{c}{Cool secondary} \\
\colhead{RA} & \colhead{DEC} &
\colhead{obs\_id} &  
\colhead{SNRU} &\colhead{source\_id} & \colhead{parallax} & \colhead{RUWE} & \colhead{AEN} & \colhead{$G$}& 
\colhead{$T_\mathrm{eff}$} & \colhead{ $\mathrm{log}\ g$} & 
\colhead{ $\mathrm{log}(n\mathrm{He}/n\mathrm{H})$} & \colhead{spclass} & \colhead{$T_\mathrm{eff}$} & \colhead{ $\mathrm{log}\ g$} & \colhead{ $\mathrm{[Fe/H]}$} & \colhead{spclass}\\
\colhead{LAMOST} & \colhead{LAMOST} & \colhead{LAMOST} &  \colhead{LAMOST} & \colhead{\textit{Gaia}} & \colhead{\textit{Gaia}} & \colhead{\textit{Gaia}} & \colhead{\textit{Gaia}(mas)} & \colhead{\textit{Gaia}(mag)} & \colhead{(\textit{K})} & \colhead{(cm\  s$^{-2}$)} & \colhead{} &  \colhead{} &  \colhead{(\textit{K})} &  \colhead{(cm\  s$^{-2}$)} &  \colhead{} 
} 
\colnumbers
\startdata
1.025383 & 23.03016 & 385811215 & 25 & 2848140084112307968 & 0.4737 & 1.155 & 0.145 & 14.87 & 36680$\pm$970 & 6.13$\pm$0.11 & -3.75$\pm$0.67 & sdB & 5997 & 4.50 & -1.00 & F5V \\
4.129504 & 31.96126 & 679405182 & 61 & 2863196315306344832 & 0.6322 & 1.151 & 0.164 & 15.37 & 29930$\pm$860 & 5.60$\pm$0.06 & -1.41$\pm$0.11 & sdB & 6918 & 4.89 & -0.77 & K1V \\
4.8358078 & 46.551897 & 282604062 & 69 & 392046852459641472 & 0.4802 & 1.487 & 0.21 & 14.25 & 27300$\pm$430 & 5.35$\pm$0.06 & -2.73$\pm$0.53 & sdB & 6998 & 5.00 & -0.25 & F2V \\
5.7646$^{*}$ $^{\dagger}$ & 24.3192361 & 163101192 & 28 & 2801217974425183104 & 0.442 & 1.249 & 0.193 & 15.44 & 35500$\pm$540 & 5.98$\pm$0.17 & -1.43$\pm$0.07 & sdOB & 4954 & 4.91 & -0.79 & K2V \\
6.473608$^{*}$ $^{\dagger}$ & 30.08673 & 360303217 & 52 & 2859002017748155776 & 0.7413 & 1.087 & 0.092 & 15.06 & 36860$\pm$160 & 5.75$\pm$0.06 & -0.84$\pm$0.04 & He-sdOB & 4931 & 4.85 & -1.07 & K3V \\
8.101786 & 39.183097 & 697009110 & 22 & 380626225180762624 & 1.2321 & 9.657 & 4.814 & 16.06 & 30020$\pm$620 & 5.57$\pm$0.03 & -2.21$\pm$0.02 & sdB & 7244 & 5.00 & -0.65 & F0V \\
10.525503$^{*}$ $^{\dagger}$ & 5.156504 & 250216131 & 62 & 2554377961881706240 & 1.091 & 3.027 & 0.669 & 12.78 & 29720$\pm$180 & 5.56$\pm$0.20 & -3.10$\pm$0.01 & sdB & 5248 & 4.98 & -0.10 & K0V \\
11.997777$^{*}$ & 3.629061 & 250209201 & 21 & 2550973079312403712 & 1.7178 & 1.322 & 0.206 & 12.35 & 44470$\pm$720 & 6.47$\pm$0.04 & -1.97$\pm$0.02 & sdO & 6471 & 4.95 & -0.51 & F7V \\
12.8622568$^{*}$ & 9.3591384 & 685611057 & 30 & 2581810261598516096 & 0.797 & 1.014 & 0.058 & 14.18 & 32860$\pm$160 & 5.85$\pm$0.22 & -2.29$\pm$0.03 & sdOB & 6251 & 4.60 & -1.00 & F5V \\
13.701168$^{*}$ $^{\dagger}$ & 6.090181 & 588011177 & 30 & 2553323049193849856 & 0.4132 & 1.21 & 0.255 & 16.03 & 27600$\pm$500 & 5.50$\pm$0.20 & -3.04$\pm$0.04 & sdB & 4753 & 4.07 & -1.87 & K3V \\
16.7959409 & 51.1731233 & 248403037 & 26 & 404392370879544576 & 3.2414 & 4.589 & 0.759 & 12.08 & 35720$\pm$300 & 5.66$\pm$0.05 & -1.32$\pm$0.09 & sdOB & 5984 & 4.26 & -0.77 & F9.5VI \\
18.3119735$^{*}$ $^{\dagger}$ & 26.4586702 & 256010167 & 70 & 306837892464925696 & 1.1299 & 1.457 & 0.235 & 12.84 & 26910$\pm$50 & 5.56$\pm$0.13 & -2.23$\pm$0.30 & sdB & 6295 & 4.96 & -0.69 & F8V \\
20.438896 & 26.94654 & 281001143 & 37 & 296407654581453824 & 0.5291 & 1.184 & 0.113 & 14.61 & 26390$\pm$140 & 5.53$\pm$0.04 & -2.01$\pm$0.03 & sdB & 5997 & 3.50 & -0.75 & G0IV \\
26.809005 & 2.629267 & 384203051 & 26 & 2512318137425806720 & 0.521 & 1.005 & 0.064 & 15.62 & 34700$\pm$330 & 6.01$\pm$0.14 & -1.51$\pm$0.14 & sdOB & 6397 & 4.80 & -0.67 & F7V \\
41.060213$^{*}$ $^{\dagger}$ & 30.123075 & 108311209 & 11 & 129720592299213056 & 0.5517 & 1.187 & 0.125 & 14.55 & 32750$\pm$1060 & 5.79$\pm$0.18 & -1.67$\pm$0.13 & sdB & 6982 & 5.00 & -0.95 & F2.5V \\
49.08386$^{*}$ $^{\dagger}$ & 0.706325 & 633209035 & 15 & 3266182479530739328 & 0.6927 & 1.045 & 0.081 & 15.10 & 32260$\pm$170 & 5.65$\pm$0.06 & -4.26$\pm$0.58 & sdB & 6456 & 4.92 & -0.75 & F7V \\
49.1125671$^{*}$ $^{\dagger}$ & 14.5351142 & 593905069 & 11 & 30166132300899584 & 0.5398 & 0.998 & 0.0 & 15.80 & 46050$\pm$3560 & 6.68$\pm$0.11 & -1.88$\pm$0.17 & sdO & 5997 & 4.61 & -0.69 & G1V \\
53.069742$^{*}$ $^{\dagger}$ & -2.550527 & 307910106 & 36 & 3261350091567233920 & 1.019 & 1.291 & 0.13 & 12.56 & 34290$\pm$250 & 5.52$\pm$0.28 & -1.94$\pm$0.01 & sdB & 6918 & 4.34 & -0.08 & F0V \\
58.278732 & 16.813495 & 280504020 & 35 & 43520323793261440 & -0.503 & 14.706 & 4.874 & 14.71 & 25520$\pm$110 & 5.72$\pm$0.07 & -3.44$\pm$0.16 & sdB & 4497 & 5.00 & -0.01 & F3IV \\
58.3071641$^{*}$ $^{\dagger}$ & 25.7560623 & 758711067 & 328 & 67303928533459968 & 2.3552 & 2.715 & 0.37 & 10.12 & 40410$\pm$540 & 5.31$\pm$0.08 & 1.06$\pm$0.10 & He-sdOB & 5495 & 2.00 & -1.00 & F6IV/V \\
60.279188$^{*}$ $^{\dagger}$ & 15.596102 & 280506023 & 45 & 40154478181532544 & 0.9774 & 0.97 & 0.0 & 14.26 & 38170$\pm$520 & 6.03$\pm$0.02 & -1.12$\pm$0.03 & sdOB & 6714 & 4.92 & -0.48 & F5V \\
60.8944383$^{*}$ $^{\dagger}$ & 26.0665106 & 264511018 & 46 & 162510331145182976 & 2.2913 & 2.039 & 0.282 & 12.27 & 31680$\pm$530 & 6.11$\pm$0.06 & -5.67$\pm$0.28 & sdB & 6456 & 4.91 & -0.25 & F7V \\
66.6441208$^{*}$ $^{\dagger}$ & 16.9241472 & 420209010 & 20 & 3313745222245617664 & 1.1816 & 1.597 & 0.145 & 13.79 & 33290$\pm$1550 & 5.87$\pm$0.29 & -3.59$\pm$0.71 & sdB & 6918 & 5.00 & -0.33 & K1V \\
74.997049$^{*}$ $^{\dagger}$ & 16.406036 & 402712100 & 11 & 3405336220170607360 & 0.5485 & 1.034 & 0.0 & 15.87 & 31900$\pm$170 & 6.70$\pm$0.04 & -1.81$\pm$0.07 & sdOB & 6501 & 4.85 & -0.93 & F6V \\
75.79921 & 42.25386 & 363101011 & 16 & 201795194412994432 & 1.1 & 0.946 & 0.135 & 14.31 & 30700$\pm$890 & 5.24$\pm$0.08 & -2.23$\pm$0.08 & sdB & 3749 & 5.00 & -2.50 & M2V \\
80.535505 & 39.628034 & 252808221 & 24 & 187949697318585856 & 0.9701 & 0.983 & 0.131 & 15.19 & 39110$\pm$660 & 5.71$\pm$0.10 & -1.02$\pm$0.04 & sdOB & 7261 & 4.99 & 0.52 & F0V \\
103.2504 & 34.471604 & 189108156 & 27 & 939393979967463424 & 0.453 & 1.177 & 0.198 & 15.58 & 34540$\pm$460 & 5.81$\pm$0.07 & -3.31$\pm$0.36 & sdB & 7227 & 5.00 & -1.46 & F0V \\
103.3741167 & 25.2871169 & 686711052 & 31 & 3381734176584838528 & 0.7634 & 1.039 & 0.164 & 12.78 & 28400$\pm$440 & 6.66$\pm$0.04 & -2.45$\pm$0.01 & sdB & 7780 & 4.88 & -0.22 & A8V \\
106.11298 & 36.912729 & 336403222 & 35 & 946065369848575872 & 0.6036 & 1.535 & 0.34 & 15.37 & 29640$\pm$450 & 5.69$\pm$0.03 & -2.92$\pm$0.14 & sdB & 4064 & 4.63 & -1.25 & K7V \\
108.903728 & 0.351749 & 778603202 & 13 & 3111129624664152064 & 1.1696 & 3.57 & 1.287 & 16.29 & 28080$\pm$110 & 5.21$\pm$0.08 & -2.66$\pm$0.19 & sdB & 4753 & 5.00 & -1.06 & K3V \\
109.54489 & 9.9644258 & 446310079 & 11 & 3156485441381866368 & 0.6282 & 5.439 & 1.04 & 14.75 & 29400$\pm$3840 & 6.92$\pm$1.44 & -0.62$\pm$1.37 & He-sdOB & 8035 & 5.00 & -0.47 & A8V \\
109.89297 & 11.856675 & 386307217 & 40 & 3163010252616980608 & 0.718 & 1.763 & 0.29 & 14.31 & 61800$\pm$650 & 5.25$\pm$0.36 & -1.41$\pm$0.08 & sdO & 5942 & 5.00 & -0.59 & G1V \\
112.2235416$^{*}$ $^{\dagger}$ & 53.7169814 & 425914149 & 24 & 983992164455024512 & 0.7676 & 1.045 & 0.068 & 13.92 & 34550$\pm$710 & 5.88$\pm$0.13 & -1.47$\pm$0.07 & sdOB & 6237 & 3.94 & -1.12 & F3V \\
112.8519 & 37.216993 & 184404173 & 27 & 898706032489528064 & 0.2873 & 1.039 & 0.031 & 15.36 & 32170$\pm$1220 & 5.48$\pm$0.12 & -4.59$\pm$1.26 & sdB & 6760 & 5.00 & -0.23 & F5V \\
114.30113$^{*}$ $^{\dagger}$ & 26.706866 & 3407169 & 11 & 872072016870129024 & 0.6281 & 1.219 & 0.168 & 15.07 & 30200$\pm$330 & 5.78$\pm$0.08 & -4.38$\pm$0.02 & sdB & 5997 & 5.00 & -1.25 & G0V \\
114.46526 & 17.03408 & 417216043 & 35 & 3169072650493941248 & 0.8258 & 1.436 & 0.209 & 14.17 & 37980$\pm$700 & 5.89$\pm$0.05 & -1.39$\pm$0.47 & sdOB & 6471 & 5.00 & -0.21 & F7V \\
115.34127$^{*}$ $^{\dagger}$ & 26.907168 & 183710065 & 36 & 869167966142655616 & 0.4764 & 1.062 & 0.131 & 15.57 & 33750$\pm$420 & 5.47$\pm$0.05 & -3.11$\pm$0.66 & sdB & 6760 & 5.00 & -0.98 & F5V \\
115.80866$^{*}$ $^{\dagger}$ & 40.564067 & 334308156 & 20 & 923503803562868608 & 0.5006 & 1.079 & 0.098 & 15.11 & 33550$\pm$1010 & 5.01$\pm$0.10 & -2.65$\pm$0.13 & sdB & 7161 & 5.00 & -0.58 & F0V \\
120.74917$^{*}$ & 41.243885 & 279212200 & 24 & 921531554580422016 & 0.5674 & 1.005 & 0.193 & 15.25 & 44880$\pm$540 & 5.08$\pm$0.13 & 2.42$\pm$0.47 & He-sdOB & 6266 & 4.89 & 0.75 & F8V \\
123.52831$^{*}$ $^{\dagger}$ & 20.317123 & 268114178 & 13 & 675664193216392704 & 0.3463 & 1.052 & 0.104 & 15.74 & 26480$\pm$180 & 5.23$\pm$0.21 & -1.95$\pm$0.04 & sdOB & 6516 & 5.00 & -0.73 & F6V \\
125.0139583$^{*}$ $^{\dagger}$ & 17.6539556 & 698903042 & 38 & 656345533398638464 & 0.408 & 0.998 & 0.0 & 15.09 & 39240$\pm$860 & 5.60$\pm$0.03 & -1.59$\pm$0.04 & sdOB & 7112 & 4.71 & -0.39 & F2V \\
126.14188$^{*}$ $^{\dagger}$ & 30.481857 & 287014148 & 14 & 708829002962898304 & 0.2979 & 1.02 & 0.067 & 15.20 & 49440$\pm$3260 & 6.09$\pm$0.22 & -1.48$\pm$0.02 & sdO & 6516 & 3.65 & -0.73 & F5V \\
126.279986$^{*}$ $^{\dagger}$ & 20.110442 & 602205226 & 34 & 663802902294594944 & 0.5254 & 1.008 & 0.06 & 14.84 & 34040$\pm$490 & 6.27$\pm$0.05 & -1.47$\pm$0.12 & sdOB & 6501 & 4.97 & -0.50 & F6V \\
126.324962$^{*}$ $^{\dagger}$ & 11.518417 & 19506063 & 13 & 601188910547673728 & 0.6931 & 1.032 & 0.077 & 14.63 & 32480$\pm$500 & 5.88$\pm$0.05 & -1.27$\pm$0.04 & sdB & 6223 & 4.52 & -0.79 & F8V \\
127.26092$^{*}$ $^{\dagger}$ & 22.776878 & 321902192 & 24 & 677934238056557312 & 0.4904 & 1.028 & 0.071 & 15.80 & 38730$\pm$460 & 6.27$\pm$0.06 & -1.47$\pm$0.13 & sdOB & 6081 & 4.62 & -0.74 & G0V \\
130.5431333$^{*}$ $^{\dagger}$ & -2.2211 & 309414071 & 28 & 3072364177559581696 & 0.8832 & 1.017 & 0.125 & 12.98 & 28860$\pm$940 & 6.26$\pm$0.03 & -3.96$\pm$0.24 & sdB & 7413 & 5.00 & -0.34 & F0V \\
130.662525 & 3.019274 & 313902052 & 78 & 3079278864452011904 & 1.7164 & 1.155 & 0.158 & 12.24 & 35970$\pm$910 & 6.58$\pm$0.09 & -1.29$\pm$0.20 & sdOB & 6456 & 4.95 & -0.05 & F6V \\
130.8190788 & 25.0614547 & 779313103 & 19 & 690626278727938560 & 1.023 & 1.013 & 0.0 & 13.44 & 30830$\pm$480 & 5.70$\pm$0.13 & -2.63$\pm$0.17 & sdB & 6137 & 4.77 & -0.36 & G0V \\
134.858699$^{*}$ $^{\dagger}$ & 7.987002 & 503806026 & 20 & 585028765383068288 & 0.3241 & 0.965 & 0.0 & 16.29 & 35080$\pm$310 & 5.68$\pm$0.04 & -1.58$\pm$0.03 & sdOB & 6109 & 5.00 & -0.89 & G0V \\
139.12066$^{*}$ $^{\dagger}$ & 24.653384 & 187512028 & 12 & 687691269876173056 & 0.4114 & 1.083 & 0.106 & 15.35 & 27680$\pm$610 & 5.11$\pm$0.05 & -2.53$\pm$0.09 & sdB & 6531 & 4.84 & -0.48 & F7V \\
150.19358$^{*}$ $^{\dagger}$ & 2.9024758 & 316603031 & 14 & 3836440886741718784 & 0.2373 & 1.058 & 0.101 & 16.36 & 37420$\pm$1700 & 5.72$\pm$0.04 & -1.25$\pm$0.15 & sdOB & 6501 & 4.47 & -0.50 & F7V \\
150.80983$^{*}$ $^{\dagger}$ & 21.856699 & 139801107 & 16 & 629594278053946752 & 0.2882 & 1.0 & 0.085 & 15.95 & 36920$\pm$3700 & 6.59$\pm$0.45 & -1.43$\pm$0.21 & sdOB & 6501 & 5.00 & -0.50 & F6V \\
150.83197$^{*}$ $^{\dagger}$ & 37.276345 & 400914083 & 42 & 802020653597763584 & 0.7726 & 1.099 & 0.131 & 14.92 & 28180$\pm$30 & 5.85$\pm$0.07 & -1.53$\pm$0.08 & sdOB & 4731 & 3.46 & -0.77 & K3V \\
154.48904$^{*}$ $^{\dagger}$ & 35.921533 & 400906249 & 23 & 753807962069995648 & 0.323 & 2.136 & 0.66 & 15.99 & 35170$\pm$450 & 6.03$\pm$0.03 & -1.39$\pm$0.23 & sdOB & 6251 & 4.89 & -0.70 & F8V \\
155.63489$^{*}$ $^{\dagger}$ & 34.415639 & 545115199 & 27 & 752309185986742912 & 0.1894 & 1.056 & 0.0 & 16.08 & 38080$\pm$400 & 5.72$\pm$0.20 & -1.59$\pm$0.08 & sdOB & 6251 & 3.36 & -0.75 & F6III \\
164.771809$^{*}$ $^{\dagger}$ & 32.105757 & 200313100 & 38 & 736957808935190656 & 0.4385 & 1.16 & 0.137 & 14.63 & 32370$\pm$510 & 5.17$\pm$0.03 & -2.25$\pm$0.32 & sdB & 6966 & 5.00 & -0.55 & F2.5V \\
165.54621$^{*}$ $^{\dagger}$ & 26.279526 & 389604033 & 27 & 3996780498461572480 & 0.3261 & 1.04 & 0.061 & 15.06 & 31100$\pm$120 & 6.78$\pm$0.07 & -1.87$\pm$0.08 & sdB & 6338 & 4.66 & -0.74 & A0V \\
166.859281$^{*}$ & 24.053095 & 385107015 & 165 & 3995351343799453056 & 3.4105 & 2.301 & 0.478 & 11.20 & 32310$\pm$180 & 6.25$\pm$0.02 & -1.77$\pm$0.02 & sdOB & 5997 & 3.99 & -0.75 & G0V \\
168.322177$^{*}$ $^{\dagger}$ & 4.220697 & 619310054 & 76 & 3812248366755810944 & 0.9561 & 1.156 & 0.168 & 14.73 & 33090$\pm$50 & 5.86$\pm$0.01 & -1.73$\pm$0.07 & sdOB & 4246 & 5.00 & -0.59 & K7V \\
170.55461$^{*}$ $^{\dagger}$ & 14.439381 & 120710031 & 18 & 3966770015776476928 & 0.2078 & 1.047 & 0.181 & 16.11 & 34100$\pm$300 & 5.67$\pm$0.02 & -3.66$\pm$0.03 & sdB & 6237 & 4.47 & -1.25 & F3V \\
172.809697$^{*}$ $^{\dagger}$ & 9.5386 & 103506220 & 23 & 3914518169503182848 & 1.0965 & 1.681 & 0.293 & 14.36 & 35050$\pm$540 & 5.96$\pm$0.02 & -1.71$\pm$0.05 & sdOB & 5432 & 3.75 & -1.69 & G8V \\
174.2646708$^{*}$ $^{\dagger}$ & 14.8445056 & 291306190 & 12 & 3972199752086976000 & 0.0895 & 3.442 & 1.275 & 16.58 & 31680$\pm$860 & 5.81$\pm$0.09 & -1.43$\pm$0.16 & sdOB & 6776 & 5.00 & -1.02 & F5V \\
180.8311583$^{*}$ $^{\dagger}$ & 9.1643472 & 815013177 & 11 & 3906031520285024512 & 0.5757 & 1.213 & 0.148 & 14.61 & 37420$\pm$1070 & 6.18$\pm$0.16 & -1.13$\pm$0.04 & sdOB & 6165 & 4.91 & -1.04 & F8V \\
183.1606958$^{*}$ $^{\dagger}$ & 42.6672917 & 149310130 & 17 & 1536809808987373824 & 0.5874 & 0.987 & 0.034 & 14.89 & 38060$\pm$190 & 5.70$\pm$0.06 & -1.24$\pm$0.09 & sdB & 6338 & 4.67 & -0.75 & A0V \\
183.2190457$^{*}$ $^{\dagger}$ & 33.9006855 & 433107100 & 71 & 4028634144887638016 & 1.2878 & 1.129 & 0.128 & 12.62 & 40600$\pm$190 & 5.76$\pm$0.06 & -1.32$\pm$0.01 & sdOB & 6516 & 4.95 & -0.46 & F6V \\
184.399697$^{*}$ $^{\dagger}$ & 37.973573 & 497901202 & 27 & 1532409357294653824 & 0.3042 & 1.125 & 0.172 & 15.52 & 27800$\pm$610 & 5.45$\pm$0.09 & -2.56$\pm$0.07 & sdB & 5997 & 3.03 & -0.78 & G0III \\
188.2903208$^{*}$ $^{\dagger}$ & 8.576175 & 220701149 & 26 & 3902809057861940480 & 0.463 & 0.901 & 0.0 & 15.49 & 36730$\pm$1280 & 6.14$\pm$0.06 & -1.01$\pm$0.07 & sdOB & 5754 & 3.50 & -0.48 & F3V \\
188.712025$^{*}$ $^{\dagger}$ & 49.789125 & 448512117 & 34 & 1544410045677548288 & 0.9806 & 2.009 & 0.349 & 13.91 & 32000$\pm$1160 & 5.71$\pm$0.11 & -2.53$\pm$0.04 & sdB & 6966 & 4.54 & -0.69 & F2V \\
192.518425$^{*}$ $^{\dagger}$ & 55.100589 & 152807229 & 48 & 1570548009752602240 & 4.0109 & 0.721 & 0.432 & 12.67 & 32160$\pm$230 & 6.06$\pm$0.07 & -0.91$\pm$0.48 & sdOB & 4753 & 5.00 & -0.18 & K3V \\
192.75618$^{*}$ $^{\dagger}$ & 37.180479 & 336710181 & 22 & 1517524890432603520 & 0.3881 & 1.067 & 0.1 & 15.53 & 50640$\pm$2860 & 6.04$\pm$0.09 & -1.54$\pm$0.03 & sdO & 6382 & 4.75 & -0.88 & F4V \\
194.0204403 & 28.1221093 & 459203172 & 26 & 1463780590268542592 &  &  & 4.673 & 13.24 & 29450$\pm$1720 & 5.46$\pm$0.09 & -2.88$\pm$0.12 & sdB & 6067 & 3.14 & -0.46 & G0III \\
195.0576423$^{*}$ $^{\dagger}$ & -2.831275 & 105604229 & 14 & 3685172413454924544 & 0.7025 & 1.019 & 0.054 & 13.96 & 30130$\pm$1410 & 5.13$\pm$0.07 & -2.38$\pm$0.08 & sdB & 6237 & 3.48 & -0.74 & F8V \\
198.387516$^{*}$ $^{\dagger}$ & 36.946957 & 563102088 & 10 & 1474359575755115136 & 0.9446 & 1.113 & 0.292 & 14.53 & 26970$\pm$1540 & 5.51$\pm$0.07 & -3.08$\pm$0.88 & sdB & 5997 & 5.00 & 0.50 & G0V? \\
199.1605083$^{*}$ $^{\dagger}$ & 3.8050667 & 645310072 & 15 & 3715866036458578688 & 0.8111 & 3.738 & 1.628 & 16.14 & 33470$\pm$510 & 5.65$\pm$0.04 & -1.58$\pm$0.13 & sdOB & 4753 & 5.00 & -1.50 & K3V \\
199.9733469 & 12.0661869 & 414507200 & 134 & 3739487119636315520 & 1.7705 & 1.849 & 0.458 & 11.26 & 41890$\pm$520 & 5.18$\pm$0.02 & -2.13$\pm$0.05 & sdO & 6745 & 4.98 & -0.43 & F5V \\
201.313488$^{*}$ $^{\dagger}$ & 38.883599 & 563108029 & 28 & 1476298186553761664 & 0.3035 & 1.069 & 0.124 & 16.22 & 37260$\pm$430 & 6.06$\pm$0.14 & -1.70$\pm$0.15 & sdOB & 5701 & 4.90 & -1.00 & G3V \\
207.73692$^{*}$ $^{\dagger}$ & 8.019409 & 451307046 & 13 & 3721767012285435520 & 0.3076 & 1.004 & 0.0 & 15.76 & 33220$\pm$2640 & 5.30$\pm$0.19 & -2.19$\pm$0.15 & sdB & 6683 & 4.50 & -0.75 & F5V \\
210.0127583$^{*}$ $^{\dagger}$ & 23.7499944 & 316208195 & 21 & 1256917132588516224 & 0.2031 & 2.808 & 0.656 & 15.14 & 44350$\pm$1480 & 5.75$\pm$0.58 & 0.27$\pm$0.15 & He-sdOB & 5199 & 4.55 & -2.25 & G8V \\
210.3218$^{*}$ $^{\dagger}$ & 27.6451083 & 332010163 & 16 & 1450983271353426048 & 0.2796 & 1.003 & 0.0 & 15.99 & 37780$\pm$1160 & 6.01$\pm$0.09 & -1.08$\pm$0.08 & sdOB & 7063 & 5.00 & -0.94 & F2V \\
210.6369292$^{*}$ $^{\dagger}$ & 32.2559694 & 230104216 & 11 & 1454622105085650688 & 0.4233 & 0.981 & 0.0 & 14.97 & 35150$\pm$210 & 5.64$\pm$0.03 & -4.01$\pm$0.25 & sdB & 6412 & 4.82 & -0.75 & F7V \\
210.821918$^{*}$ $^{\dagger}$ & 22.172126 & 573612151 & 29 & 1256509935329189120 &  &  & 16.67 & 15.91 & 35340$\pm$870 & 5.67$\pm$0.14 & -1.01$\pm$0.02 & sdOB & 6011 & 5.00 & -1.23 & G0V \\
212.317154$^{*}$ $^{\dagger}$ & 38.475605 & 242115019 & 27 & 1485199008058786688 & 0.4356 & 1.027 & 0.069 & 16.04 & 43170$\pm$370 & 6.00$\pm$0.12 & -1.48$\pm$0.18 & sdO & 5997 & 5.00 & -0.75 & G0V \\
215.27003$^{*}$ $^{\dagger}$ & 17.608 & 425408223 & 20 & 1232778626111037184 & 0.3108 & 1.079 & 0.157 & 16.27 & 33870$\pm$120 & 5.78$\pm$0.02 & -1.38$\pm$0.09 & sdOB & 5997 & 4.99 & -1.04 & G0V \\
215.8780875$^{*}$ $^{\dagger}$ & 14.8089694 & 302514109 & 15 & 1228629618983512064 & 0.2106 & 0.987 & 0.117 & 15.26 & 34520$\pm$100 & 5.14$\pm$0.08 & -1.93$\pm$0.27 & sdB & 6745 & 5.00 & -1.21 & F5V \\
215.8983417$^{*}$ $^{\dagger}$ & 34.2387222 & 438302106 & 46 & 1479772643298162176 & 0.6344 & 1.017 & 0.0 & 14.70 & 40060$\pm$1800 & 6.39$\pm$0.29 & -1.43$\pm$0.02 & sdOB & 6251 & 5.00 & -0.25 & G0V \\
222.0906375$^{*}$ $^{\dagger}$ & 31.5184417 & 660103039 & 23 & 1283389867897035392 & 0.2348 & 1.03 & 0.0 & 16.24 & 38820$\pm$40 & 5.09$\pm$0.91 & -1.39$\pm$0.09 & sdOB & 6501 & 4.97 & -0.49 & F6V \\
227.77946$^{*}$ $^{\dagger}$ & 17.537572 & 581401069 & 17 & 1211503282272546432 & 0.4243 & 0.999 & 0.0 & 16.15 & 32270$\pm$880 & 5.67$\pm$0.11 & -2.60$\pm$0.10 & sdB & 6441 & 4.39 & -0.67 & F6V \\
229.055053$^{*}$ & 27.648553 & 558307059 & 23 & 1271587607005565440 & 0.1436 & 1.037 & 0.211 & 16.48 & 53820$\pm$1770 & 5.70$\pm$0.18 & -1.58$\pm$0.19 & sdO & 5834 & 4.68 & -1.66 & F8V \\
229.3097958$^{*}$ $^{\dagger}$ & 3.1742833 & 723007009 & 18 & 4421940758497428096 & 0.8424 & 1.343 & 0.187 & 13.91 & 32190$\pm$100 & 5.85$\pm$0.03 & -1.73$\pm$0.02 & sdOB & 5984 & 4.63 & -0.83 & G1V \\
229.535314$^{*}$ & 4.178841 & 237508148 & 14 & 1155791990165521920 & 0.5414 & 1.125 & 0.18 & 15.31 & 37620$\pm$110 & 6.13$\pm$0.32 & -1.79$\pm$0.05 & sdOB & 5956 & 3.97 & -1.22 & G1V \\
230.611347$^{*}$ $^{\dagger}$ & 24.726743 & 581904203 & 12 & 1270016817203702400 & 0.3453 & 0.981 & 0.0 & 15.60 & 30920$\pm$700 & 5.45$\pm$0.09 & -2.24$\pm$0.12 & sdB & 6652 & 4.53 & -0.34 & F5V \\
231.01266$^{*}$ $^{\dagger}$ & 1.572694 & 431411183 & 19 & 4420741294390728448 & 0.9773 & 1.564 & 0.215 & 13.87 & 23720$\pm$690 & 5.28$\pm$0.07 & -3.15$\pm$0.14 & sdB & 5069 & 5.00 & 0.32 & K1V \\
231.4727042$^{*}$ $^{\dagger}$ & 43.6909778 & 745010247 & 25 & 1394116942282089344 & 0.7262 & 1.187 & 0.176 & 15.05 & 36850$\pm$100 & 5.75$\pm$0.16 & -0.51$\pm$0.01 & He-sdOB & 5000 & 5.00 & 0.00 & K2V \\
233.013331$^{*}$ $^{\dagger}$ & 42.962491 & 573411193 & 21 & 1391344214475144832 & 0.5505 & 1.142 & 0.164 & 15.23 & 35800$\pm$2590 & 6.29$\pm$0.33 & -1.34$\pm$0.42 & sdOB & 6683 & 4.86 & -0.71 & F5V \\
234.530673$^{*}$ $^{\dagger}$ & 27.6954506 & 221010099 & 10 & 1224346643237436288 & 0.7733 & 1.019 & 0.013 & 13.72 & 40460$\pm$4510 & 5.12$\pm$0.54 & -1.30$\pm$0.09 & sdOB & 6561 & 4.62 & -0.44 & F6V \\
238.3402865$^{*}$ $^{\dagger}$ & 25.4037151 & 662002168 & 17 & 1220195948186053760 & 1.5392 & 2.778 & 0.53 & 14.19 & 23550$\pm$450 & 5.68$\pm$0.03 & -4.06$\pm$0.60 & sdB & 4753 & 5.00 & -0.50 & K3V \\
240.486021 & 6.314934 & 568315063 & 37 & 4450708930484104960 & 0.7211 & 1.404 & 0.207 & 14.82 & 27250$\pm$20 & 5.45$\pm$0.13 & -2.63$\pm$0.06 & sdB & 5188 & 5.00 & -0.56 & K0V \\
240.537772$^{*}$ & 7.419737 & 568316229 & 32 & 4451111935856224000 & 0.7051 & 0.976 & 0.093 & 14.46 & 73400$\pm$19720 & 5.96$\pm$0.13 & -1.74$\pm$0.18 & sdO & 5847 & 5.00 & -0.41 & G2V \\
241.54936$^{*}$ $^{\dagger}$ & 3.8714259 & 744001168 & 25 & 4413410266254905344 & 0.3737 & 0.986 & 0.0 & 15.48 & 34510$\pm$560 & 5.96$\pm$0.07 & -1.23$\pm$0.12 & sdOB & 6918 & 4.84 & -0.61 & F2V \\
243.1328155$^{*}$ $^{\dagger}$ & 57.2476258 & 812903110 & 11 & 1621996774452667648 & 0.1664 & 0.976 & 0.0 & 17.48 & 33720$\pm$1890 & 5.58$\pm$0.12 & -6.66$\pm$0.01 & sdB & 6295 & 4.60 & -0.99 & A0V \\
244.526825$^{*}$ $^{\dagger}$ & 21.6903639 & 461106067 & 25 & 1202447086032118784 & 0.6055 & 1.046 & 0.073 & 14.80 & 31960$\pm$420 & 6.71$\pm$0.17 & -1.55$\pm$0.05 & sdOB & 6237 & 5.00 & -1.01 & F8V \\
252.59188$^{*}$ $^{\dagger}$ & 31.463808 & 347313144 & 12 & 1313140659676082560 & 0.5871 & 1.144 & 0.184 & 15.82 & 37680$\pm$390 & 6.10$\pm$0.04 & -0.19$\pm$0.03 & He-sdOB & 4742 & 4.48 & -0.39 & K3V \\
252.791963$^{*}$ $^{\dagger}$ & 8.059091 & 331602037 & 23 & 4442343330624541568 & 0.6845 & 1.262 & 0.1 & 13.71 & 40490$\pm$210 & 5.80$\pm$0.11 & -1.20$\pm$0.19 & sdOB & 6194 & 4.87 & -0.31 & F8V \\
254.674322$^{*}$ & 41.521005 & 566911146 & 25 & 1354579844177083904 & 0.4121 & 0.933 & 0.0 & 15.99 & 33360$\pm$410 & 5.67$\pm$0.05 & -1.92$\pm$0.05 & sdOB & 5861 & 4.21 & -1.14 & G1.5V \\
258.9182625$^{*}$ $^{\dagger}$ & 34.3821528 & 576806137 & 10 & 1334991769651067904 & 0.362 & 1.081 & 0.115 & 15.49 & 29940$\pm$300 & 5.45$\pm$0.11 & -1.79$\pm$0.08 & sdB & 6501 & 5.00 & -0.75 & F6V \\
259.938993 & 51.869609 & 580916012 & 22 & 1415898405068162176 &  &  & 8.596 & 13.80 & 32290$\pm$850 & 5.37$\pm$0.16 & -3.66$\pm$0.29 & sdB & 6095 & 5.00 & -0.66 & G0V \\
261.6163772$^{*}$ $^{\dagger}$ & 37.155398 & 141504052 & 13 & 1337240374008902784 & 0.9327 & 1.19 & 0.086 & 13.31 & 41310$\pm$960 & 6.26$\pm$0.15 & -1.86$\pm$0.17 & sdO & 6934 & 4.85 & -0.50 & F2V \\
261.9515653$^{*}$ & 16.7492625 & 342803178 & 62 & 4550114402362108416 & 1.1289 & 2.847 & 0.524 & 13.48 & 27550$\pm$90 & 5.43$\pm$0.01 & -3.17$\pm$0.20 & sdB & 4497 & 5.00 & -0.86 & K7V \\
268.5154$^{*}$ $^{\dagger}$ & 53.6932417 & 155403168 & 14 & 1369049829516110336 & 0.5823 & 1.036 & 0.035 & 15.26 & 48160$\pm$8070 & 5.26$\pm$0.16 & -1.62$\pm$0.18 & sdO & 6237 & 5.00 & 0.05 & G0V \\
272.61257 & 14.261412 & 746402012 & 25 & 4497397527098931712 & 0.4661 & 1.05 & 0.21 & 16.29 & 25790$\pm$210 & 5.80$\pm$0.04 & -2.44$\pm$0.01 & sdB & 4753 & 4.45 & -0.23 & K2V \\
273.35417 & 6.8877511 & 742403249 & 22 & 4478006677464647040 & 0.5729 & 1.379 & 0.36 & 15.95 & 46110$\pm$1850 & 5.14$\pm$0.20 & 0.76$\pm$1.05 & He-sdOB & 4753 & 5.00 & 0.50 & K3V \\
278.15735 & 11.778036 & 746709096 & 10 & 4481070569714541056 & 0.4538 & 0.978 & 0.0 & 17.19 & 36290$\pm$720 & 5.72$\pm$0.09 & -1.49$\pm$0.23 & sdB & 6652 & 4.81 & -0.50 & F5V \\
284.946255$^{*}$ $^{\dagger}$ & 57.6758294 & 346813058 & 19 & 2153548180479601280 & 0.3662 & 1.001 & 0.064 & 14.81 & 40410$\pm$1080 & 5.88$\pm$0.16 & -2.74$\pm$0.13 & sdO & 6745 & 4.97 & -0.75 & F5V \\
287.5008566$^{*}$ $^{\dagger}$ & 46.6734565 & 354401198 & 35 & 2130430570550718464 & 0.498 & 1.007 & 0.065 & 14.47 & 44360$\pm$5230 & 5.76$\pm$0.20 & -1.84$\pm$0.47 & sdO & 6251 & 4.57 & -1.01 & F5V \\
289.0894514 & 48.4702516 & 353802136 & 60 & 2132462777275978624 & 0.5795 & 1.002 & 0.07 & 14.39 & 30460$\pm$700 & 5.56$\pm$0.21 & -2.38$\pm$0.05 & sdB & 6714 & 4.80 & -0.79 & F5V \\
290.0299828 & 52.0936831 & 353811024 & 56 & 2139296001523913728 & 0.726 & 0.986 & 0.0 & 14.39 & 33890$\pm$100 & 6.22$\pm$0.07 & -1.65$\pm$0.03 & sdOB & 6471 & 4.82 & -0.69 & F7V \\
305.11349$^{*}$ & 7.0704421 & 587205114 & 60 & 4249848840652401792 & 0.869 & 1.325 & 0.164 & 14.10 & 74200$\pm$1810 & 5.44$\pm$0.08 & 2.34$\pm$0.19 & He-sdO & 6966 & 5.00 & -0.42 & F2.5V \\
319.4259042$^{*}$ $^{\dagger}$ & -0.1055306 & 677910180 & 30 & 2689440797710883584 & 0.3705 & 1.087 & 0.086 & 15.02 & 36630$\pm$1640 & 5.83$\pm$0.13 & -1.20$\pm$0.06 & sdOB & 6382 & 4.77 & -0.75 & F7V \\
320.259476$^{*}$ $^{\dagger}$ & 12.847195 & 260814117 & 18 & 1746937544892092416 & 0.7124 & 1.265 & 0.121 & 13.49 & 48560$\pm$1710 & 5.60$\pm$0.21 & -2.63$\pm$0.12 & sdO & 6295 & 4.10 & 0.03 & F8V \\
320.6041 & 17.862396 & 593101087 & 43 & 1785228484006179456 & 0.6792 & 1.112 & 0.082 & 14.33 & 56750$\pm$3940 & 6.46$\pm$0.08 & -1.31$\pm$0.14 & sdO & 6501 & 5.00 & -0.54 & F6V \\
321.10134$^{*}$ & 15.105552 & 685407021 & 87 & 1783474797319468032 & 1.5707 & 1.502 & 0.248 & 12.95 & 39090$\pm$170 & 5.69$\pm$0.02 & -1.17$\pm$0.20 & sdOB & 6531 & 4.95 & -0.46 & F6V \\
321.22878$^{*}$ & 15.984361 & 592411060 & 54 & 1783691847786812288 & 0.8228 & 1.575 & 0.244 & 14.40 & 29930$\pm$60 & 6.30$\pm$0.77 & -2.04$\pm$0.39 & sdB & 6165 & 4.46 & -0.98 & F5V \\
321.743738$^{*}$ $^{\dagger}$ & 7.34571 & 372505162 & 32 & 1739298997095482240 & 0.9628 & 1.161 & 0.097 & 14.03 & 36680$\pm$70 & 6.19$\pm$0.01 & -1.64$\pm$0.17 & sdOB & 6251 & 4.95 & -0.73 & F8V \\
321.870759 & 31.3114233 & 368209056 & 49 & 1853416896629422464 & 0.9232 & 2.126 & 0.36 & 14.43 & 27760$\pm$70 & 5.26$\pm$0.04 & -2.62$\pm$0.04 & sdB & 4092 & 4.69 & -1.25 & K7V \\
322.275089$^{*}$ $^{\dagger}$ & 0.75258 & 677909003 & 47 & 2688012605121741312 & 0.2962 & 1.027 & 0.115 & 15.50 & 37230$\pm$440 & 5.76$\pm$0.03 & -1.24$\pm$0.18 & sdOB & 4753 & 4.50 & -0.75 & K2V \\
332.161998 & 14.819503 & 378813161 & 24 & 1774431039302375552 & 0.3556 & 1.041 & 0.091 & 15.79 & 47750$\pm$830 & 6.22$\pm$0.04 & -1.72$\pm$0.03 & sdO & 6237 & 4.94 & -0.75 & F8V \\
334.627458$^{*}$ & 18.802447 & 75311209 & 13 & 1777173805417611776 & 1.1962 & 2.117 & 0.291 & 13.84 & 31910$\pm$370 & 5.83$\pm$0.05 & -1.74$\pm$0.04 & sdOB & 4508 & 5.00 & -1.49 & K4V \\
343.32667$^{*}$ $^{\dagger}$ & 21.898433 & 490613065 & 26 & 2835948493025229952 & 0.8726 & 0.967 & 0.008 & 13.06 & 34510$\pm$230 & 5.39$\pm$0.08 & -2.56$\pm$0.26 & sdB & 6698 & 4.40 & -0.30 & F5V \\
350.122232$^{*}$ $^{\dagger}$ & 28.494027 & 249901087 & 52 & 2845145731696691968 & 1.3136 & 0.956 & 0.101 & 11.87 & 26200$\pm$480 & 6.11$\pm$1.25 & -1.93$\pm$0.52 & sdOB & 6714 & 4.42 & -0.29 & F5IV \\
352.02525$^{*}$ $^{\dagger}$ & 50.621368 & 689003056 & 19 & 1943261255910215552 & 0.2068 & 1.131 & 0.212 & 16.70 & 33640$\pm$640 & 5.74$\pm$0.07 & -6.99$\pm$0.09 & sdB & 6745 & 5.00 & -0.50 & F5V \\
354.63865$^{*}$ $^{\dagger}$ & 47.495592 & 678715215 & 30 & 1940032157758205824 & -0.0189 & 1.082 & 0.148 & 16.24 & 26600$\pm$230 & 6.58$\pm$0.13 & -5.14$\pm$0.17 & sdB & 7533 & 4.71 & -0.32 & A8V \\
\enddata
\end{deluxetable*}
\end{longrotatetable}

Table 1 gives the main properties of 131 hot subdwarf stars identified with composite spectra. From columns (1) to (4), the table presents right ascension (RA), declination (DEC), LAMOST\_obsid and signal to noise ratio in LAMOST \textit{u} band (SNRU), respectively. 
Columns (5) to (9) give the source\_id, parallaxes, RUWE values, astrometric excess noise (AEN) and magnitudes in the \textit{G} band of \textit{Gaia} EDR3. 
Columns (10) to (13) give effective temperatures ($T_{\rm eff}$), gravity ($\log{g}$), helium abundance ($\log(n{\rm He}/n{\rm H})$) and  
spectral classification for the hot subdwarf primary. While columns (14) to (17) give the effective temperatures, gravity, metallicity ([Fe/H]) and spectral classification for the companion stars based on the fitting results. 

In Table 2, the main properties of 91 hot subdwarf stars identified with single-lined spectra were presented. 
Columns (1) to (4) give the right ascension (RA), declination (DEC), LAMOST\_obsid and SNRU, respectively. 
Columns (5) to (6)  give the source\_id,  and  magnitudes in \textit{G} band of \textit{Gaia} EDR3, while columns (7) to (10) give the detailed atmospheric parameters (i.e., $T_{\rm eff}$, $\log{g}$, $\log(n{\rm He}/n{\rm H})$ and their uncertainties) and spectral classification.
Objects with RA labeled by $^{*}$ both in Table 1 and 2 are common stars in the catalog of \citet{2020A&A...635A.193G} and also have reported atmospheric parameters, while objects with RA labeled by $^{*}$ and $^{\dagger}$ are common stars in the catalog of \citet{2020A&A...635A.193G} but without atmospheric parameters there. 
\begin{deluxetable*}{llllllllll}
\tablenum{2}
\tablecaption{91 hot subdwarf stars identified by single-lined spectra.}
\tablewidth{100pt}
\tabletypesize{\tiny}
\tablehead{
\colhead{RA} & \colhead{DEC} & 
\colhead{obs\_id} &  
\colhead{SNRU} &\colhead{source\_id} & \colhead{$G$}  &
\colhead{$T_\mathrm{eff}$} & \colhead{ $\mathrm{log}\ g$} & 
\colhead{ $\mathrm{log}(n\mathrm{He}/n\mathrm{H})$} & \colhead{spclass}\\
\colhead{LAMOST} & \colhead{LAMOST} & \colhead{LAMOST} &  \colhead{LAMOST} & \colhead{\textit{Gaia}} & \colhead{\textit{Gaia}(mag)} & \colhead{(\textit{K})} & \colhead{(cm\  s$^{-2}$)} & \colhead{} &  \colhead{} 
}
\colnumbers
\startdata
7.4749346 & 50.541949 & 594011103 & 30 & 391755859836391680 & 16.18 & 29210$\pm$560 & 5.49$\pm$0.13 & -2.92$\pm$0.37 & sdB \\
17.2632275$^{*}$ $^{\dagger}$ & 52.7300771 & 248315093 & 20 & 404959035980962944 & 15.49 & 32250$\pm$460 & 5.41$\pm$0.04 & -2.60$\pm$0.10 & sdB \\
29.0050281 & 40.0561321 & 264312040 & 122 & 344155939885410816 & 11.43 & 28380$\pm$330 & 5.31$\pm$0.03 & -2.62$\pm$0.04 & sdB \\
32.881148$^{*}$ $^{\dagger}$ & 39.287209 & 613410220 & 31 & 332606704106366848 & 15.54 & 35370$\pm$990 & 5.94$\pm$0.10 & -1.64$\pm$0.03 & sdOB \\
33.933972$^{*}$ $^{\dagger}$ & 33.668954 & 601516009 & 108 & 326009496898928896 & 13.72 & 27990$\pm$190 & 5.44$\pm$0.02 & -2.81$\pm$0.03 & sdB \\
36.524742 & 55.982399 & 678103210 & 35 & 457462812156715008 & 16.06 & 37640$\pm$460 & 5.39$\pm$0.07 & -3.25$\pm$0.40 & sdB \\
37.082246 & 36.948326 & 778503239 & 15 & 328366815470365952 & 16.84 & 28180$\pm$280 & 5.26$\pm$0.04 & -1.95$\pm$0.02 & sdB \\
37.34398 & 55.681828 & 678104014 & 13 & 457250400255056768 & 17.12 & 28040$\pm$220 & 5.22$\pm$0.09 & -2.91$\pm$0.04 & sdB \\
45.366714 & 30.260161 & 198207185 & 29 & 123108816565129344 & 14.97 & 60550$\pm$4720 & 6.02$\pm$0.12 & -2.25$\pm$0.16 & sdO \\
46.533112$^{*}$ & 38.393252 & 473710095 & 11 & 142548697540212352 & 16.80 & 36800$\pm$1990 & 6.11$\pm$0.20 & -1.93$\pm$0.11 & sdOB \\
49.405835$^{*}$ & 37.486865 & 255501173 & 37 & 235323736345986176 & 15.30 & 28980$\pm$220 & 5.48$\pm$0.02 & -2.90$\pm$0.02 & sdB \\
54.679217 & 30.388769 & 188007106 & 15 & 120354952254440960 & 16.57 & 32920$\pm$540 & 5.58$\pm$0.09 & -1.48$\pm$0.05 & sdB \\
57.507775$^{*}$ & 32.74251 & 301301092 & 11 & 217062978512976000 & 17.24 & 31530$\pm$250 & 5.97$\pm$0.54 & -2.19$\pm$0.04 & sdB \\
61.6862792 & 22.7204186 & 796803063 & 33 & 53566385439973376 & 14.37 & 28980$\pm$170 & 5.55$\pm$0.02 & -2.99$\pm$0.06 & sdB \\
65.145303$^{*}$ & 1.3446668 & 777816028 & 84 & 3255780171819962496 & 12.30 & 44790$\pm$320 & 5.33$\pm$0.07 & 1.91$\pm$0.35 & He-sdOB \\
73.813501$^{*}$ & 13.091574 & 402801174 & 13 & 3296244226946647168 & 17.15 & 29420$\pm$610 & 5.37$\pm$0.09 & -2.68$\pm$0.04 & sdB \\
77.99079 & 23.93592 & 757003099 & 21 & 3418841422712575616 & 16.72 & 41070$\pm$520 & 5.44$\pm$0.01 & 1.44$\pm$0.04 & He-sdOB \\
80.752383$^{*}$ $^{\dagger}$ & 18.457499 & 616504175 & 23 & 3400623129219934336 & 16.56 & 40200$\pm$670 & 5.40$\pm$0.06 & 2.47$\pm$0.03 & He-sdB \\
82.637651$^{*}$ & 11.345517 & 606308098 & 15 & 3340622921426651648 & 16.71 & 28850$\pm$340 & 5.80$\pm$0.16 & -2.88$\pm$0.23 & sdB \\
88.065822 & 43.88204 & 714502122 & 13 & 193503502346470784 & 16.24 & 27510$\pm$390 & 5.52$\pm$0.03 & -2.58$\pm$0.09 & sdB \\
88.9259708$^{*}$ & 64.1363 & 614505205 & 22 & 287018271958898560 & 13.24 & 53290$\pm$1940 & 5.49$\pm$0.07 & -2.16$\pm$0.15 & sdO \\
110.41193 & 5.7395204 & 769308212 & 14 & 3140865744837849216 & 14.86 & 42220$\pm$620 & 5.50$\pm$0.05 & -0.21$\pm$0.07 & He-sdOB \\
114.9939542$^{*}$ $^{\dagger}$ & 17.9763889 & 725308189 & 23 & 671353867118310528 & 14.92 & 31330$\pm$180 & 5.46$\pm$0.03 & -3.27$\pm$0.12 & sdB \\
115.35463 & 12.838747 & 688807112 & 11 & 3163891442466252928 & 17.25 & 39300$\pm$560 & 5.64$\pm$0.12 & -0.20$\pm$0.09 & He-sdOB \\
118.700879 & 1.770195 & 779707047 & 78 & 3087687074686316416 & 14.24 & 34720$\pm$250 & 5.91$\pm$0.05 & -0.74$\pm$0.04 & He-sdOB \\
120.18431 & 18.067203 & 712912143 & 14 & 669097841055847296 & 16.04 & 29360$\pm$320 & 5.31$\pm$0.05 & -1.56$\pm$0.12 & sdB \\
127.00848$^{*}$ $^{\dagger}$ & 40.669153 & 811303149 & 12 & 914500864915422848 & 17.92 & 43160$\pm$1370 & 5.58$\pm$0.07 & -1.92$\pm$0.49 & sdO \\
131.48705$^{*}$ $^{\dagger}$ & 13.8700278 & 792102231 & 14 & 609252591386688256 & 17.43 & 24530$\pm$270 & 5.58$\pm$0.14 & -3.94$\pm$0.59 & sdB \\
131.5497792$^{*}$ $^{\dagger}$ & 24.4201722 & 779213036 & 12 & 690346315579638656 & 15.58 & 28930$\pm$390 & 5.48$\pm$0.05 & -3.08$\pm$0.03 & sdB \\
133.0738375$^{*}$ & 21.2768861 & 700706139 & 10 & 685761455170072192 & 16.30 & 29580$\pm$360 & 5.58$\pm$0.06 & -6.10$\pm$0.60 & sdB \\
135.7208206$^{*}$ & 7.592742 & 781214052 & 15 & 584194373496664064 & 17.30 & 42750$\pm$360 & 6.29$\pm$0.08 & 1.26$\pm$0.06 & He-sdOB \\
150.1573529 & 18.2564037 & 778710002 & 14 & 626203208100096896 & 16.87 & 28750$\pm$850 & 5.38$\pm$0.21 & -2.50$\pm$0.15 & sdB \\
150.17257$^{*}$ & 36.752124 & 803414182 & 13 & 795998387893409152 & 16.97 & 27280$\pm$300 & 5.47$\pm$0.10 & -2.73$\pm$0.07 & sdB \\
153.175889$^{*}$ & 48.82705 & 613703104 & 25 & 822945768622050048 & 16.45 & 29910$\pm$1040 & 5.50$\pm$0.15 & -2.44$\pm$0.12 & sdB \\
156.16429$^{*}$ $^{\dagger}$ & 38.654983 & 811404140 & 11 & 755712414993199744 & 16.78 & 30830$\pm$400 & 5.67$\pm$0.10 & -2.16$\pm$0.07 & sdB \\
166.1875917$^{*}$ & 9.4255167 & 660505073 & 31 & 3867259957150344960 & 16.27 & 36720$\pm$280 & 5.20$\pm$0.06 & -2.30$\pm$0.09 & sdOB \\
170.261064$^{*}$ $^{\dagger}$ & 5.816939 & 619315126 & 101 & 3813980166288451072 & 14.21 & 29870$\pm$30 & 5.58$\pm$0.04 & -3.07$\pm$0.40 & sdB \\
171.3625$^{*}$ & 11.4839972 & 779809104 & 10 & 3915826137367906816 & 17.34 & 34350$\pm$490 & 6.01$\pm$0.05 & -0.96$\pm$0.06 & He-sdOB \\
178.49506$^{*}$ & 35.658008 & 633010131 & 35 & 4031612751951454080 & 16.47 & 30110$\pm$270 & 5.59$\pm$0.05 & -3.25$\pm$0.22 & sdB \\
180.96765$^{*}$ & 23.895365 & 629804097 & 30 & 4002664328779500928 & 16.99 & 40020$\pm$720 & 5.75$\pm$0.10 & -0.51$\pm$0.08 & He-sdOB \\
190.5072255$^{*}$ & 43.6731474 & 803515195 & 22 & 1540495440685138432 & 16.92 & 35540$\pm$670 & 5.89$\pm$0.10 & -0.34$\pm$0.05 & He-sdOB \\
192.6661291$^{*}$ $^{\dagger}$ & 75.8833605 & 644104043 & 29 & 1692116238728076544 & 15.36 & 72360$\pm$3110 & 5.47$\pm$0.10 & 0.64$\pm$0.11 & He-sdO \\
193.8111 & 37.541553 & 795403228 & 31 & 1517595224818622976 & 15.74 & 56060$\pm$1490 & 7.20$\pm$0.08 & -2.10$\pm$0.26 & sdO \\
196.3399572 & 38.9833266 & 795409037 & 24 & 1523599314219554432 & 16.79 & 40770$\pm$540 & 6.52$\pm$0.15 & -2.29$\pm$0.05 & sdO \\
201.9527625$^{*}$ & 9.9139972 & 783505160 & 68 & 3732055520543728256 & 13.94 & 34640$\pm$650 & 5.83$\pm$0.04 & -1.82$\pm$0.03 & sdOB \\
202.089054$^{*}$ $^{\dagger}$ & 5.148918 & 645309049 & 40 & 3717793239823667200 & 14.35 & 42240$\pm$1790 & 5.39$\pm$0.12 & 1.92$\pm$0.20 & He-sdOB \\
203.92918$^{*}$ & 33.099549 & 545205055 & 10 & 1469167024717489152 & 18.37 & 43500$\pm$920 & 5.44$\pm$0.11 & 0.93$\pm$0.15 & He-sdOB \\
207.719937$^{*}$ $^{\dagger}$ & 36.700612 & 805407025 & 71 & 1495299843425776000 & 13.36 & 66760$\pm$2380 & 6.41$\pm$0.07 & -1.07$\pm$0.22 & sdO \\
220.058975$^{*}$ & 23.1266528 & 644814213 & 57 & 1242599704488616704 & 14.91 & 29930$\pm$220 & 5.59$\pm$0.02 & -4.37$\pm$0.12 & sdB \\
220.6141583$^{*}$ & -1.545975 & 441415005 & 62 & 3649025693813906304 & 13.85 & 40230$\pm$430 & 5.44$\pm$0.06 & -2.96$\pm$0.11 & sdO \\
222.1817542$^{*}$ $^{\dagger}$ & 8.5894722 & 803609201 & 20 & 1173616997796784384 & 16.17 & 51940$\pm$2520 & 6.46$\pm$0.39 & -2.49$\pm$0.26 & sdO \\
222.390187$^{*}$ & 24.726785 & 644811215 & 35 & 1266451513309671296 & 15.68 & 35140$\pm$460 & 5.90$\pm$0.06 & -1.79$\pm$0.03 & sdOB \\
228.6353042$^{*}$ & 24.1775917 & 239010244 & 46 & 1264108144792924160 & 13.18 & 31100$\pm$130 & 5.62$\pm$0.04 & -1.89$\pm$0.03 & sdB \\
230.5968583$^{*}$ $^{\dagger}$ & 13.3179361 & 820615101 & 50 & 1170796406874734464 & 15.36 & 41690$\pm$620 & 5.14$\pm$0.09 & 2.04$\pm$0.07 & He-sdOB \\
231.53285$^{*}$ & 0.2779916 & 820909174 & 12 & 4417304530279961088 & 16.82 & 57590$\pm$1470 & 5.77$\pm$0.07 & -0.98$\pm$0.08 & He-sdO \\
234.6305$^{*}$ $^{\dagger}$ & 29.958647 & 346201172 & 10 & 1273157297289530368 & 17.28 & 21650$\pm$540 & 5.02$\pm$0.11 & -1.98$\pm$0.08 & sdB \\
236.537625$^{*}$ $^{\dagger}$ & 25.1279833 & 822913146 & 63 & 1222767813259931776 & 14.06 & 49120$\pm$470 & 5.52$\pm$0.05 & 0.79$\pm$0.09 & He-sdOB \\
248.455582$^{*}$ $^{\dagger}$ & 26.549607 & 820308185 & 45 & 1304445275047376640 & 15.41 & 28490$\pm$130 & 5.55$\pm$0.02 & -2.81$\pm$0.02 & sdB \\
248.5808667$^{*}$ $^{\dagger}$ & 34.7125861 & 823308245 & 20 & 1327508145619011840 & 17.11 & 34730$\pm$780 & 5.96$\pm$0.05 & -0.50$\pm$0.04 & He-sdOB \\
259.072204$^{*}$ & 55.5795951 & 660208123 & 12 & 1420619570557801472 & 17.17 & 31230$\pm$430 & 5.59$\pm$0.07 & -3.00$\pm$0.15 & sdB \\
259.34208$^{*}$ & 58.099688 & 660212084 & 16 & 1433879818228293632 & 16.74 & 35870$\pm$800 & 5.36$\pm$0.12 & -1.92$\pm$0.08 & sdOB \\
260.79634 & 17.969148 & 742215078 & 34 & 4553448465214622848 & 14.65 & 32580$\pm$250 & 5.74$\pm$0.04 & -1.59$\pm$0.03 & sdOB \\
267.866027$^{*}$ $^{\dagger}$ & 39.386402 & 821009068 & 24 & 4610982060405219712 & 15.98 & 36190$\pm$490 & 5.74$\pm$0.04 & -1.54$\pm$0.04 & sdOB \\
268.0235794 & 40.3014294 & 821012227 & 49 & 1344525738054881152 & 14.42 & 21500$\pm$100 & 5.05$\pm$0.03 & -2.10$\pm$0.04 & sdB \\
269.93357 & 4.2326256 & 743810217 & 12 & 4469765146325951360 & 16.31 & 28870$\pm$560 & 5.38$\pm$0.11 & -2.38$\pm$0.07 & sdB \\
272.94594 & 6.5129634 & 742410048 & 16 & 4471987980878277120 & 16.75 & 36030$\pm$720 & 5.98$\pm$0.11 & -3.30$\pm$0.54 & sdB \\
273.88709 & 6.7067732 & 742403036 & 17 & 4477802580630416384 & 16.49 & 26110$\pm$730 & 5.63$\pm$0.13 & -3.98$\pm$0.08 & sdB \\
\enddata
\end{deluxetable*}

\begin{deluxetable*}{llllllllll}
\tablenum{2}
\tablecaption{continued}
\tablewidth{100pt}
\tablehead{
\colhead{RA} & \colhead{DEC} & 
\colhead{obs\_id} &  
\colhead{SNRU} &\colhead{source\_id} & \colhead{$G$}  &
\colhead{$T_\mathrm{eff}$} & \colhead{ $\mathrm{log}\ g$} & 
\colhead{ $\mathrm{log}(n\mathrm{He}/n\mathrm{H})$} & \colhead{spclass}\\
\colhead{LAMOST} & \colhead{LAMOST} & \colhead{LAMOST} & \colhead{LAMOST} & \colhead{\textit{Gaia}} & \colhead{\textit{Gaia}(mag)} & \colhead{(\textit{K})} & \colhead{(cm\  s$^{-2}$)} & \colhead{} &  \colhead{} 
}
\colnumbers
\startdata
276.01834 & 5.1066987 & 742407233 & 15 & 4284459778774532224 & 15.93 & 60850$\pm$2560 & 5.35$\pm$0.11 & -0.48$\pm$0.11 & He-sdO \\
276.04245 & 32.374372 & 743607178 & 23 & 4591910996864556032 & 15.48 & 27290$\pm$730 & 5.47$\pm$0.11 & -2.07$\pm$0.14 & sdB \\
276.65441$^{*}$ $^{\dagger}$ & 34.623952 & 743612201 & 20 & 2095370374556456192 & 15.65 & 33740$\pm$760 & 5.95$\pm$0.07 & -0.86$\pm$0.05 & He-sdOB \\
277.25692 & 30.525699 & 821113141 & 27 & 4588505431396750336 & 15.26 & 26900$\pm$220 & 5.45$\pm$0.03 & -2.70$\pm$0.01 & sdB \\
277.323911 & 11.532653 & 820714096 & 22 & 4483962525804601856 & 15.39 & 34630$\pm$1520 & 5.62$\pm$0.13 & -3.90$\pm$0.04 & sdB \\
277.42603 & 11.074758 & 746715115 & 17 & 4480928797123755904 & 16.66 & 46990$\pm$1790 & 5.15$\pm$0.23 & 2.24$\pm$0.14 & He-sdOB \\
277.8296 & 9.3204009 & 746708185 & 32 & 4480111417612829952 & 16.04 & 27790$\pm$1890 & 5.39$\pm$0.27 & -2.49$\pm$0.09 & sdB \\
277.84971 & 9.503416 & 746708180 & 20 & 4480150415917049728 & 16.31 & 30190$\pm$420 & 5.52$\pm$0.04 & -2.67$\pm$0.03 & sdB \\
280.23555 & 7.6993656 & 743114037 & 20 & 4286765592098890624 & 16.27 & 27760$\pm$400 & 5.52$\pm$0.04 & -2.48$\pm$0.03 & sdB \\
306.59602 & 9.3368847 & 587312198 & 12 & 1752454585005332992 & 17.25 & 43600$\pm$1530 & 5.34$\pm$0.07 & 0.67$\pm$0.05 & He-sdOB \\
313.32814$^{*}$ $^{\dagger}$ & 35.931016 & 680410248 & 19 & 1869904107861540864 & 15.65 & 30670$\pm$270 & 5.74$\pm$0.18 & -2.50$\pm$0.13 & sdB \\
316.56508$^{*}$ $^{\dagger}$ & 34.320724 & 680407016 & 12 & 1867119354139385984 & 15.89 & 47210$\pm$1320 & 6.06$\pm$0.12 & -0.19$\pm$0.21 & He-sdOB \\
320.5687 & 12.515551 & 592501071 & 21 & 1746866317154525696 & 16.59 & 30620$\pm$480 & 5.52$\pm$0.09 & -2.43$\pm$0.08 & sdB \\
321.01139 & 37.647009 & 593805115 & 20 & 1964027697669073408 & 16.99 & 48650$\pm$1720 & 5.51$\pm$0.10 & -2.40$\pm$0.21 & sdO \\
326.23313 & 32.669941 & 767514200 & 11 & 1946116172210792704 & 17.27 & 62980$\pm$2060 & 5.87$\pm$0.09 & -2.41$\pm$0.17 & sdO \\
327.05073 & 32.458548 & 767503052 & 13 & 1946414341722524672 & 17.55 & 36220$\pm$600 & 5.93$\pm$0.06 & -1.47$\pm$0.06 & sdOB \\
327.32986 & 32.942349 & 767503123 & 19 & 1946462930693095680 & 15.40 & 28170$\pm$650 & 5.51$\pm$0.07 & -2.48$\pm$0.15 & sdB \\
327.54146 & 30.236853 & 767502103 & 43 & 1897581731796584960 & 15.74 & 31050$\pm$160 & 5.71$\pm$0.05 & -2.56$\pm$0.04 & sdB \\
327.62997 & 32.364481 & 767503170 & 36 & 1946231483493079936 & 16.54 & 46030$\pm$330 & 5.43$\pm$0.03 & 3.84$\pm$0.14 & He-sdB \\
327.7554 & 30.729705 & 767502069 & 27 & 1897622589823405184 & 14.27 & 46290$\pm$750 & 5.43$\pm$0.14 & 1.76$\pm$0.16 & He-sdOB \\
329.42852 & 34.160007 & 767512146 & 13 & 1948026230067409152 & 16.34 & 28630$\pm$290 & 5.35$\pm$0.09 & -2.22$\pm$0.02 & sdB \\
338.8813792$^{*}$ & 14.5662944 & 354514045 & 38 & 2732982626402453888 & 13.99 & 34020$\pm$340 & 5.13$\pm$0.05 & -2.85$\pm$0.09 & sdB \\
340.7985417 & -0.0780167 & 756314177 & 18 & 2653544977175287936 & 15.63 & 30030$\pm$180 & 5.78$\pm$0.12 & -3.04$\pm$0.06 & sdB \\
344.3473878 & -6.0464972 & 756505144 & 16 & 2611711201142881536 & 17.04 & 34050$\pm$200 & 5.83$\pm$0.06 & -1.50$\pm$0.05 & sdOB \\
\enddata
\end{deluxetable*}

\begin{figure}
    \centering
    \includegraphics[width=100mm]{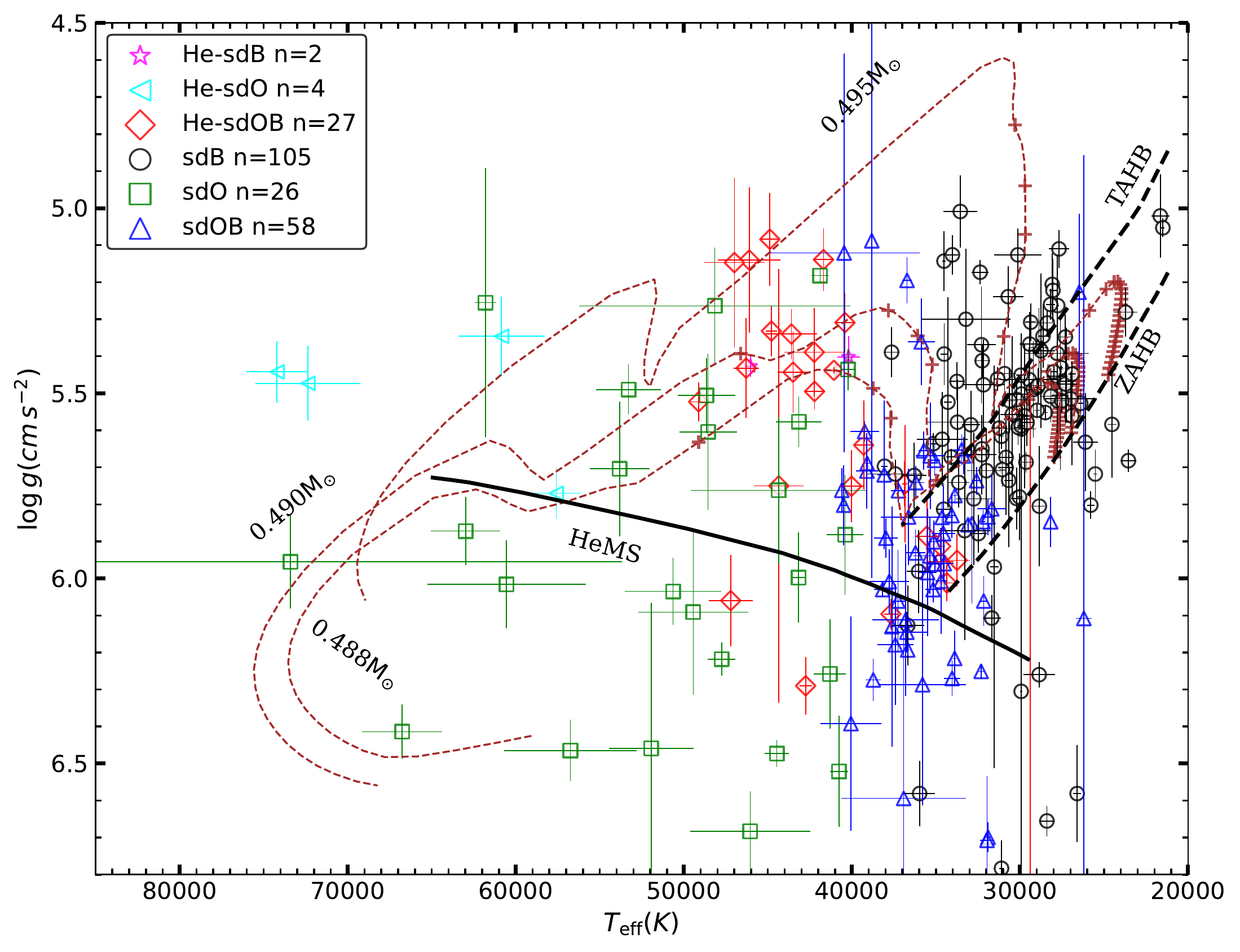}
    \caption{$T_{\rm eff}$-$\log{g}$ diagram for 222 hot subdwarf stars identified in this study. The markers and number counts for different types of hot subdwarfs are shown in the upper-left box. The zero-age horizontal branch (ZAHB) and terminal-age horizontal branch (TAHB) sequences with [Fe/H]= -1.48 from \citet{1993ApJ...419..596D} are denoted by dashed lines. The He-MS from \citet{1971AcA....21....1P} is marked by black solid line. Three evolution tracks for hot HB stars from \citet{1993ApJ...419..596D} are presented by brown dotted curves, for which the masses from top to bottom are 0.495, 0.490, and 0.488 $M_{\odot}$, respectively.}
    \label{fig4}
\end{figure}

\begin{figure}
    \centering
    \includegraphics[width=100mm]{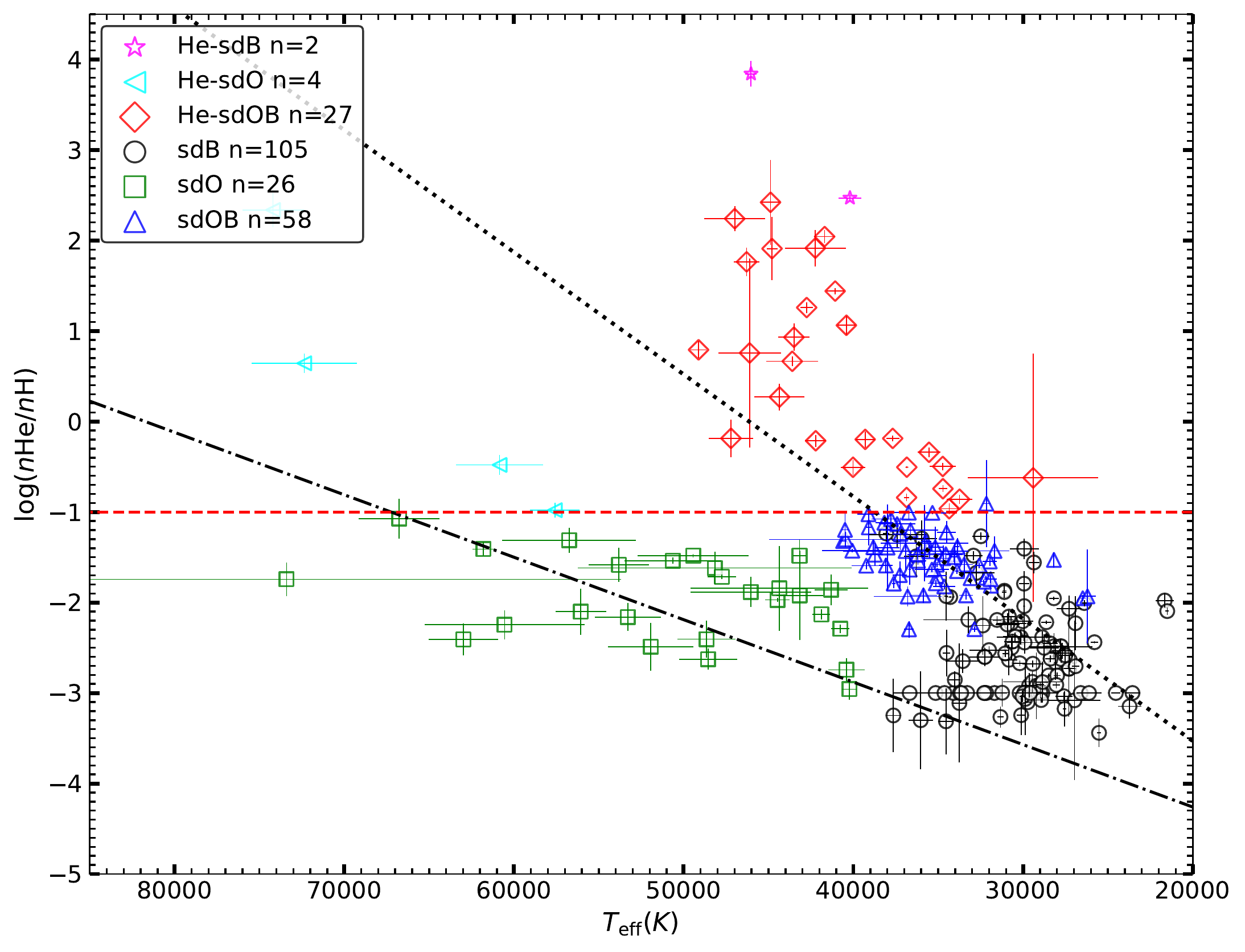}
    \caption{$T_{\rm eff}$-$\mathrm{log}(n\mathrm{He}/n\mathrm{H})$ diagram for 222 hot subdwarf identified in this study. The black dotted line and dot-dashed line are the linear regression lines fitted for the two He sequences by \citet{2003A&A...400..939E} and \citet{2012MNRAS.427.2180N}, respectively.}
    \label{fig5}
\end{figure}

\begin{figure}
    \centering
    \includegraphics[width=100mm]{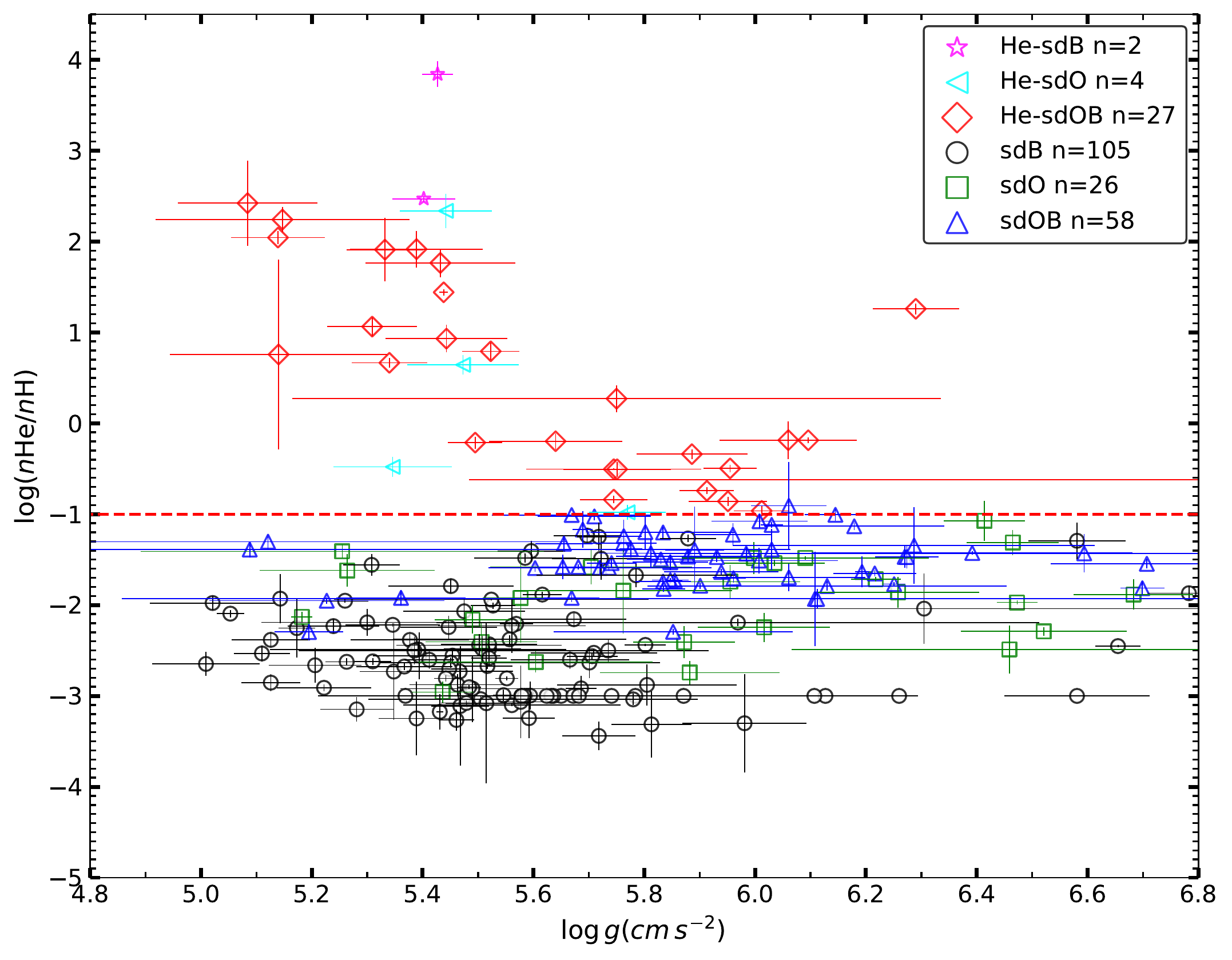}
    \caption{  $\log{g}$-$\mathrm{log}(n\mathrm{He}/n\mathrm{H})$ diagram for the 222 hot subdwarfs identified in this study. The red horizontal dashed line denotes the solar value of He abundance (e.g., $\mathrm{log}(n\mathrm{He}/n\mathrm{H})$ = -1).}
    \label{fig6}
\end{figure}

\subsection{Parameter diagrams}
The relationships between atmospheric parameters of our 222 samples, e.g., $T_\mathrm{eff}$ and $\mathrm{log}\ g$, $T_\mathrm{eff}$ and $\mathrm{log}(n\mathrm{He}/n\mathrm{H})$, 
$\mathrm{log}\ g$ and $\mathrm{log}(n\mathrm{He}/n\mathrm{H})$, are shown in Fig \ref{fig4} to \ref{fig6},  respectively. Hot subdwarf stars are labeled by different symbols (see legend boxes) in each figure based on their spectral classification. The uncertainties of the parameter values are represented by thin-solid short lines with the same color as symbols.   

The relationships between the parameters presented in Fig 4 to 6 are very similar to our previous works (see \citealt{2018ApJ...868...70L,2019ApJ...881..135L,2020ApJ...889..117L} for detailed discussion). For example, two He sequences (i.e., a  He-rich sequence and a He-poor sequence) are shown very clearly in $T_\mathrm{eff}$ - $\mathrm{log}(n\mathrm{He}/n\mathrm{H})$ diagram (see Fig \ref{fig5}), which also have been identified previously by several authors \citep{2003A&A...400..939E,2012MNRAS.427.2180N,2013A&A...557A.122G,2016ApJ...818..202L,2019ApJ...881....7L}. Moreover, He-sdOB stars (red open diamonds) are split into two subgroups near at $\mathrm{log}(n\mathrm{He}/n\mathrm{H})$ = 0, which is also presented in our previous work (see Section 5 in  \citealt{2019ApJ...881..135L} for detailed discussion). 

\section{discussion and summary} 
\subsection{Two-color diagram}
\begin{figure}
    \centering
    \includegraphics[width=100mm]{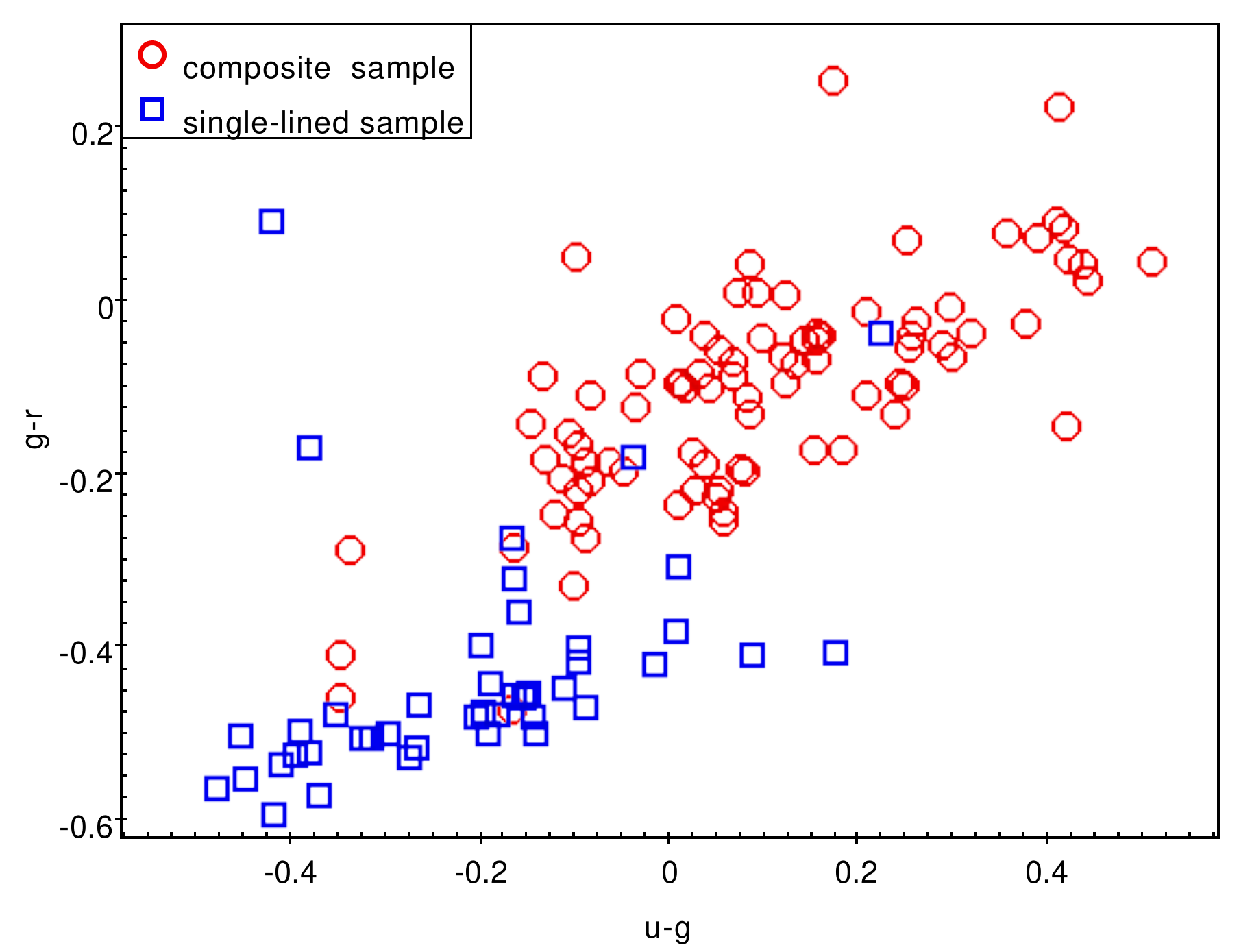}
    \caption{Two-color diagram for hot subdwarfs identified in this study. Only objects having \textit{u}, \textit{g} and \textit{r} photometry from SDSS DR12 are plotted.} Red open circles are hot subdwarf stars identified with composite spectra, while blue open squares are hot subdwarf stars identified with single-lined spectra.
    \label{fig7}
\end{figure}

Fig \ref{fig7} presents the two-color diagram for both composite and single-lined hot subdwarf stars. 
The \textit{u}, \textit{g} and \textit{r} magnitudes were obtained by crossing-matching our sample with SDSS DR12. 
The figure clearly shows that composite hot subdwarf stars (red open circles) are well separated from single-lined hot subdwarf stars (blue open squares) in the two-color diagram. 
These results are consistent with the ones described in Fig 5 of \citet{2017A&A...600A..50G} and Fig 3 of \citet{2020A&A...635A.193G}. 
Note that we selected composite spectra manually based on the presence of obvious Mg I triplet lines at 5170 \AA \ and/or Ca II triplet lines at 8650 \AA  \ in the spectra, while \citet{2020A&A...635A.193G} used color criterion in the two-color diagram (e.g., see Table 1 in their study).

\subsection{Two newly discovered wide sdB binaries.}
The third data release of the \textit{Gaia} mission, \textit{Gaia} DR3, became available to public on 13 June 2022 \citep{2022arXiv220800211G}. 
It not only contains astrometric and broad-band photometric information for more than 1.8 billion objects, which have already been published in \textit{Gaia} EDR3 \citep{2021A&A...649A...1G}, but also presents a large variety of new data products, such as: mean BP, RP and RVS spectra, a much larger radial velocity survey, a large number of variable sources and astrophysical parameters, etc. 
Especially, \textit{Gaia} DR3 also contains a non-single star catalog, in which orbital information and trend parameters were solved for more than 800\,000 astrometric, spectroscopic, and eclipsing binaries. 
This catalog provides great convenience to study the binary nature of our composite sample. 

We cross-matched our sample of 131 composite spectra with the \textit{Gaia} DR3 non-single star catalog, and found two common objects (i.e., LAMOST obs\_id = 282604062 and 342803178, named as sdB\_b1 and sdB\_b2 hereafter, respectively). Their information are listed in Table 3.
Based on our decomposition results in Table 1, sdB\_b1 consists of a sdB primary with $T_\mathrm{eff}=27300\pm430$ K,  $\log\mathrm{g}=5.36\pm0.06$ $\mathrm{cm/s^{-2}}$ and a F2V type cool secondary. 
While sdB\_b2 consists of a sdB primary with $T_\mathrm{eff}=27550\pm90$ K,  $\log\mathrm{g}=5.43\pm0.01$ and a K7V type cool secondary. 
Based on the data from \textit{Gaia} DR3, both binary systems have very long orbital periods, i.e., ${\rm P}=884\pm25$ and ${\rm P}=821\pm43$ days, respectively (see Table 3). 
However, the sdB\_b1 system has a much higher eccentricity (e.g., ${\rm e}=0.5\pm0.09$) than sdB\_b2 (e.g., ${\rm e}=0.15\pm0.04$). 

\begin{deluxetable*}{cllllllccll}
\tablenum{3}
\tablecaption{Two wide sdB binaries identified in this study with solved orbital parameters from {\it Gaia} DR3. }
\tablewidth{100pt}
\tablehead{
\colhead{Name} & \colhead{RA} & \colhead{DEC} &  \colhead{obs\_id} &
\colhead{source\_id} & \colhead{parallax} & \colhead{RUWE} & \colhead{AEN} & \colhead{$G$}  &
\colhead{Period} & \colhead{Eccentricity} \\
\colhead{This study} & \colhead{LAMOST} & \colhead{LAMOST} & \colhead{LAMOST} &  \colhead{\textit{Gaia}} & \colhead{\textit{Gaia}} & \colhead{\textit{Gaia}} & \colhead{\textit{Gaia}(mas)} & \colhead{\textit{Gaia}(mag)} & \colhead{\textit{Gaia}(days)} & \colhead{Gaia}
}
\colnumbers
\startdata
sdB\_b1 & 4.8358078 & 46.551897 & 282604062 & 392046852459641472 & 0.4802 & 1.487 & 0.21 & 14.25 & 884$\pm$25 & 0.50$\pm$0.09 \\
sdB\_b2 & 261.9515653 & 16.7492625 & 342803178 & 4550114402362108416 & 1.1289 & 2.847 & 0.524 & 13.48 & 821$\pm$43 & 0.15$\pm$0.04 \\
\enddata
\end{deluxetable*}

\begin{figure}
    \centering
    \includegraphics[width=160mm]{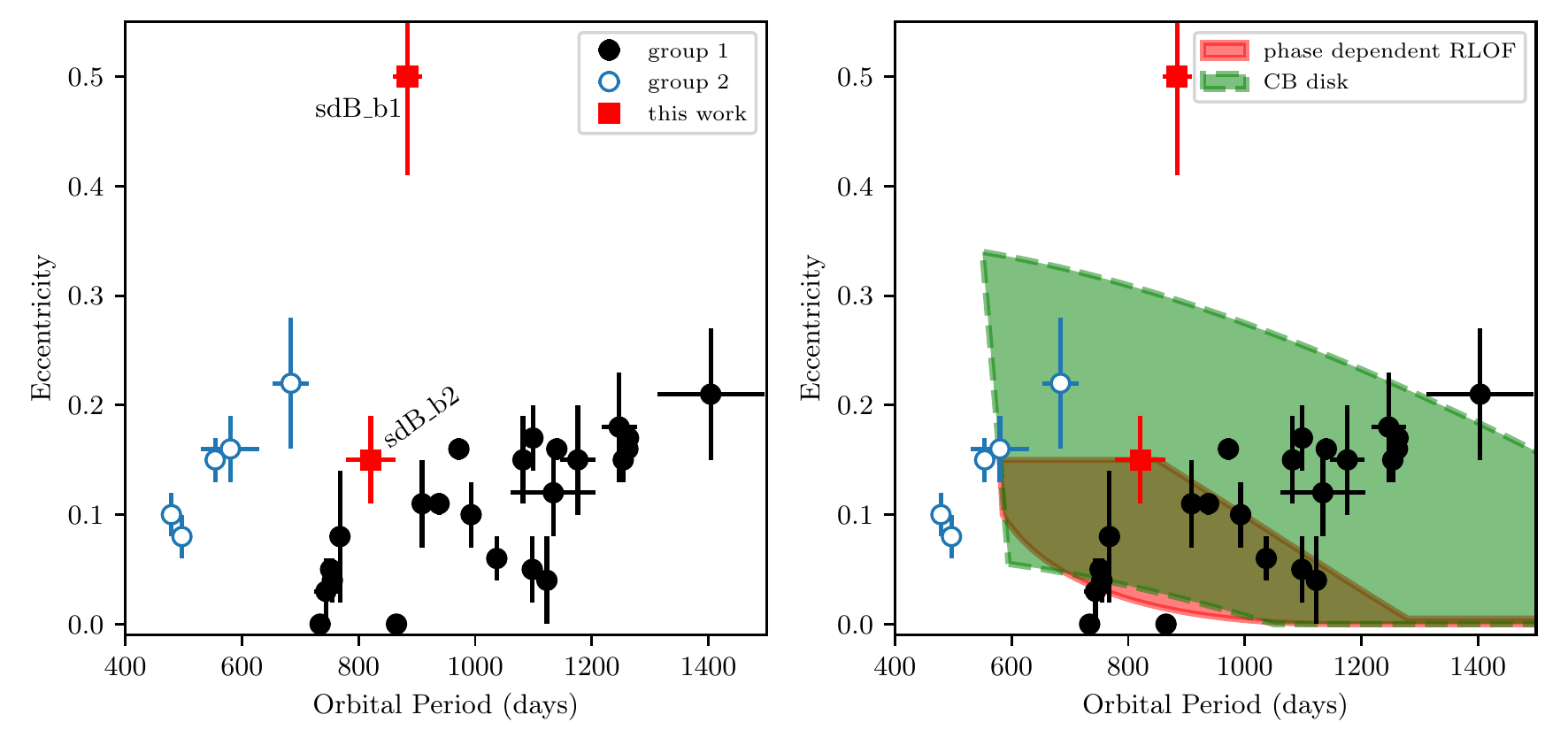}
    \caption{Period-eccentricity plane for wide sdB binaries. Left panel: Known wide sdB binaries are denoted by blue open and black filled circles, while the two newly discovered sdB binaries are shown by red squares. Right panel: same as the left panel, but added prediction area  of two physical processes from \citet{2015A&A...579A..49V} and \citet{2018MNRAS.474..433D}, red for phase-dependent RLOF, while green for CB disk process.}
    \label{fig8}
\end{figure}

So far, only 23 long-period composite sdB binaries were reported with solved orbital parameters \citep{2017A&A...605A.109V, 2018MNRAS.473..693V, 2019MNRAS.482.4592V, 2020A&A...641A.163V, 2021A&A...653A...3N, 2022A&A...658A.122M}. 
In Fig \ref{fig8}, we plot known long-orbital period sdB binaries together with the two new discoveries in the period-eccentricity plane. 
As discussed in other studies \citep{2017A&A...605A.109V, 2020A&A...641A.163V, 2021A&A...653A...3N}, there are two distinct groups among these wide sdB binary systems. 
This distinction is visible both in the period-eccentricity and the period-mass ratio plane \citep{2019MNRAS.482.4592V}. The main group (e.g., black filled circles, named Group 1 hereafter) contains most of the binary systems, and all of them present orbital periods longer than 700 days. While the minor group (e.g., blue open circle, named Group 2 hereafter) has only five binary systems, and their orbital periods are shorter than 700 days. 

Both two groups show an obvious tendency for higher eccentricities at longer orbital periods, which has not been well explained until now. 
The left panel of Fig \ref{fig8} clearly shows, that both of the two new sdB binaries discovered in this study (red filled squares) belong to Group 1 according to their longer orbital periods. 
However, sdB\_b1 presents a very high eccentricity (i.e., e=0.5) that is not only much higher than sdB\_b2 (i.e., e=0.15), but also much higher than all other wide sdB binary systems. Therefore, this result places it far away from both Groups 1 and 2. The very high eccentricity demonstrates a different formation or evolution history of this binary, and it needs more detailed and dedicated study in the future. 
On the other hand, sdB\_b2 is consistent with the members of Group 1 in the period-eccentricity plane, which suggests a similar formation history as the main group members. 

According to the results from detailed binary population synthesis \citep{2002MNRAS.336..449H, 2003MNRAS.341..669H, 2013MNRAS.434..186C}, wide sdB binaries are formed from stable RLOF channel with final orbital periods between 400-1600 days. 
Nevertheless, these binary systems are predicted with circular orbits before the onset of RLOF, while most of the observed wide sdB binaries present eccentric orbits. 
There must be some other, yet unclear mechanisms preventing orbital circularization during binary interactions. 

\citet{2015A&A...579A..49V} studied the effects of three processes in triggering eccentric orbits during RLOF.
They found that phase-dependent mass loss during RLOF and interaction of the binary with a circumbinary (CB) disk could produce eccentric sdB binary systems during RLOF. \citet{2018MNRAS.474..433D} extended these models to a wider range of CB-disk properties. 
In the right panel of Fig \ref{fig8}, the region predicted by phase-dependent RLOF (red area) and CB disk (green area) were added into the period-eccentricity plane. 
One can see that, most of the binaries from Group 1 (including sdB\_b2 discovered in this study) and parts of sdB binaries from Group 2 could be predicted by the models when the two physical processes were considered together. 
However, there are also several binary systems with shorter orbital periods in Group 2 and higher eccentricities in Group 1 out of the model predictions. 
Especially, when sdB\_b1 is considered, which has the highest eccentricity (i.e., e=0.5) among Groups 1 and 2, the situation becomes even more complicated. 
Since this binary also presents a much higher eccentricity than the maximum value from the model prediction (e.g., e $\approx$ 0.35 at a period of about 550 days,  predicted by the CB disk process). 
As discussed in \citet{2015A&A...579A..49V} and \citet{2021A&A...653A...3N}, sdB binaries present higher eccentricity with longer orbital periods in both two groups, while the two physical processes in \citet{2015A&A...579A..49V} predicted an opposite trend. Furthermore, \citet{2020A&A...642A.234O} investigated the effect of a CB-disk on the eccentricity of wide post-AGB binaries and found that the Lindbladt resonances between a CB disk and the inner binary are not capable of predicting the observed eccentricities in post-AGB systems.

In the case of the high eccentricity of sdB\_b1, it is possible that this system was a triple system where the inner binary merged to produce the current sdB component. The cool companion would then not be involved in the original interaction phase, which could explain the high eccentricity. A detailed study of the abundances of the cool companion as performed by \citet{2022A&A...658A.122M} can provide more insight into this problem. Further observational and theoretical work is needed to uncover the nature of the period-eccentricity distribution for wide sdB binaries. 

\begin{figure}
    \centering
    \includegraphics[width=100mm]{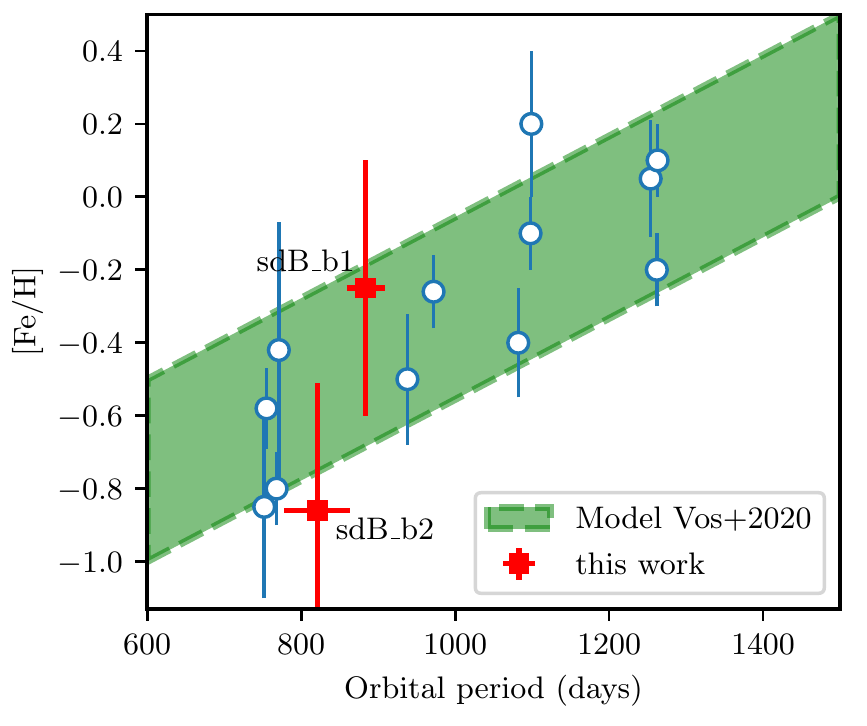}
    \caption{The period-metallicity plane for wide sdB binaries. Known wide sdB binaries are denoted by blue open circles, while the two newly discovered sdB binaries are shown by red squares. An error of 0.35 in [Fe/H] was assumed for the two new binaries. In green shade, the predicted area of \citet{2020A&A...641A.163V} based on a galactic evolution model coupled with a deterministic binary interaction model.}
    \label{fig9}
\end{figure}

\citet{2020A&A...641A.163V} investigated the expected wide sdB population using a deterministic binary interaction model including L2 and L3 mass loss coupled with a galactic evolution model. The aim was to explain the period-mass ratio relation found in wide sdB binaries \citep{2019MNRAS.482.4592V}. This model did not only explain the period-mass ratio relation, but it also predicted a period-metallicity relation originating from the metallicity evolution history in our galaxy. Fig\,\ref{fig9} shows the predicted period-metallicity area from the model together with all known observed systems. Both sdB\_b1 and sdB\_b2 fit well with these predictions, further supporting the validity of the interaction model proposed by \citet{2020A&A...641A.163V}. The relevance of this is wider than just sdB binaries, as similar models and approaches are now being used to study the observed populations of other post interaction systems as for example RR Lyr stars \citep{2022arXiv220804332B} and blue large amplitude pulsators \citep{2022A&A...XiongForthcomming}.

\begin{figure}
    \centering
    \includegraphics[width=100mm]{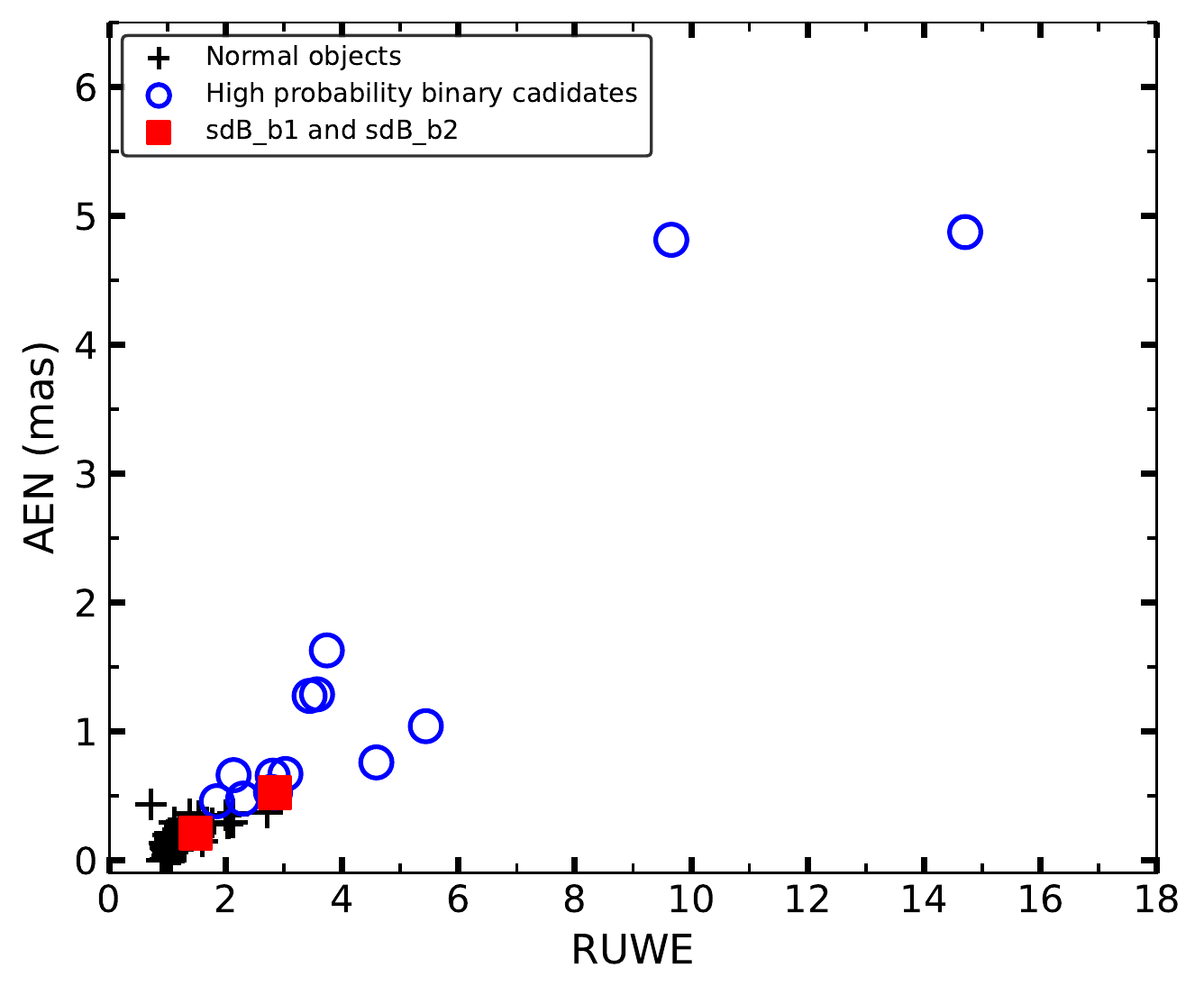}
    \caption{RUWE-AEN plane for our composite hot subdwarf sample. Objects with AEN $\geq$ 0.4 and RUWE $\geq$ 1.4 are represented by blue open circles, while normal sample are denoted by black pluses. The two nwely discovered sdB binaries are labeled by red filled squares. }
    \label{fig10}
\end{figure}

\subsection{Hints for binarity in composite spectrum hot subdwarf stars.}
A high astrometric excess noise (AEN) is a potential signature of physical binary systems. 
It is defined as the excess uncertainty that must be added in quadrature to obtain a statistically acceptable astrometric solution \citep{2018A&A...616A...1G,2012A&A...538A..78L}.
Higher than expected AEN values might appear in binary systems with unresolved companions. 
This method was first adopted by \citet{2022MNRAS.510.3885G} to search for X-ray binaries. 
On the other hand, the Renormalised Unit Weight Error (RUWE) from \textit{Gaia} data is also considered as a signature for binarity if its value is significantly greater than 1.0 \citep{2018A&A...616A...2L}. 
\citet{2022arXiv220908971U} have combined AEN with RUWE to search for low-mass red giants (RG) binaries, which are also the potential progenitors of sdB stars. 
They found that more than 90\% of the known RG binaries have AEN $\geq$ 0.4, while 93\% of known RG binaries have RUWE $\geq$ 1.4. 
The two values were applied as one of the criteria to search for RG binary candidates in their study. 

To provide more hints for the binary nature of our composite hot subdwarf stars, we plotted them in the RUWE-AEN plane. 
As shown in Fig \ref{fig10}, there are 15 composite sdB stars located in the region with AEN $\geq$ 0.4 and RUWE $\geq$ 1.4, which were labeled as high probability binary candidates (e.g., blue open circles). 
These systems deserve priority follow-up observations in the future. 
The two wide sdB binaries identified in this study (e.g., red filled squares) are also presented in the figure. 
One can see that, sdB\_b2 (e.g., AEN=0.524, RUWE=2.847) is well in the high  probability binary candidates group. 
While sdB\_b1 is out of this group due to a smaller AEN value of 0.21. 
But it still presents a large value of RUWE = 1.487, which is also a sign to consider the system as a binary candidate. 
Based on the results, some hot subdwarf binaries may be hidden in the normal group (e.g., black pluses).
In contrast to the RG samples of \citet{2022arXiv220908971U} the RUWE-AEN distribution is less sensitive for binaries in the case of hot subdwarfs as our samples have less luminous stars at larger distances, leading to larger AEN noise and RUWE errors.
Yet, according to Fig \ref{fig10}, there are 15 high-probability binary candidates in our sample. 

\begin{figure}
    \centering 
    \includegraphics[width=170mm]{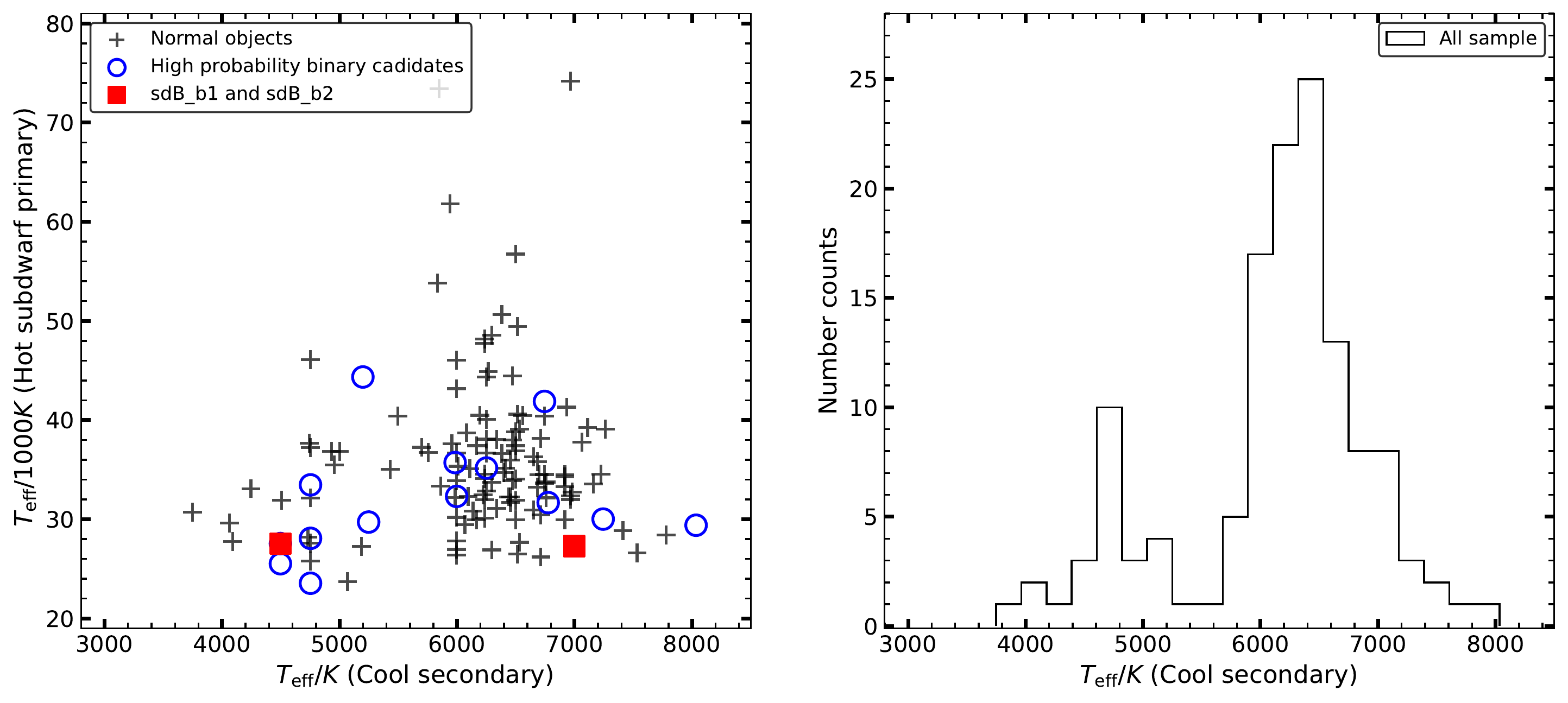}
    \caption{Left panel: $T_\mathrm{eff}$-$T_\mathrm{eff}$ plane of hot subdwarf primaries vs cool companions for our composite sample. Different labels denote the same groups as in Fig \ref{fig10}. Right panel: Histogram of $T_\mathrm{eff}$ distribution for cool companions. Note that both two panels show a distinct gap around at 5500 K among cool companions, see detailed discussions in the context.}
    \label{fig11}
\end{figure}

Fig \ref{fig11} presents the relationship of effective temperature ($T_\mathrm{eff}$) distribution between hot subdwarf primaries and cool secondaries for our composite sample. The left panel of Fig \ref{fig11} clearly shows that, both high probability binary candidates (e.g., blue open circles) and normal objects (e.g., black pluses) defined by RUWE-AEN plane in Fig \ref{fig10} were well separated by a gap among cool secondaries around at 5500 K. To show this feature more clearly, we plotted a histogram of $T_\mathrm{eff}$ distribution for all cool secondaries in the right panel. There is indeed a distinct gap around at 5500 K among cool companions. Based on our results of decomposition processes (see Sect 3 and Table 1), most of cool companions in the composite-spectra systems were F, G and K type stars. Therefore, the gap shown in Fig \ref{fig11} seems to be from the temperatures differences between FG and K type stars. 

Interestingly, results from binary population synthesis (BPS) of \citet{2003MNRAS.341..669H} also predicted a similar gap between the temperature of FG and K type companions for sdB stars produced from the first stable RLOF channel (see panel b and e of Fig 15 in their study). They concluded that the gap was caused by the Hertzsprung gap, since the companions could accrete more masses when the progenitor of sdB stars were in the Hertzsprung gap than the companions with sdB progenitors on the first giant branch (FGB), and tend to be more massive and of earlier spectral type (see detailed discussion in Sect 7.5 of \citealt{2003MNRAS.341..669H}). Note that, the hot subdwarf stars with composite spectra presented here were just binary candidates (except for the two newly discovered sdB binary systems). Some of them could be not real binaries, and their companions may be a background or foreground FG or K stars. Hence, we could not obtain a deeper conclusion on this field before the binary nature of the sample are confirmed.

\citet{2022arXiv220209608G} studied radial velocity (RV) variability for a sample of 646 hot subdwarfs with multi-epoch radial velocities from SDSS and LAMOST. 
They found that only a small fraction (e.g., $\approx$ 3\%, see Section 3 and Fig 2 in their study) of He-rich hot subdwarf stars (i.e., $\mathrm{log}(n\mathrm{He}/n\mathrm{H}) \geqslant$ -1) show RV variability, while a much higher fraction of He-poor hot subdwarf stars (i.e., $\mathrm{log}(n\mathrm{He}/n\mathrm{H}) <$-1) presents RV variability (e.g., $\approx$ 30\%). 
This striking result led the authors to the conclusion that He-rich and He-poor hot subdwarf stars are evolutionarily not related, and He-rich hot subdwarf stars likely form through the binary merger channel. Considering the results of \citet{2020A&A...642A.180P}, single-star evolution has little contribution to the formation of hot subdwarf stars. 

Note that, all the stars analyzed by \citet{2022arXiv220209608G} were single-lined hot subdwarf stars, while composite hot subdwarfs were not included. 
We cross-matched our 131 composite hot subdwarfs identified with LAMOST DR8 LRS Multiple Epoch Catalog, in which 7,912,959 sources with 2 times or more multiple observations were collected, and found 61 common records. 
Unfortunately, among these 61 common objects, only two sources have completed RV curves from LAMOST, and the RV variability is low. 
The lack of RV values for the composite systems prevents us from deriving deeper conclusions in this field. 
In near future, we plan to use the multi-epoch observed spectra to obtain RV variability of our composite sample, analyze the evolutionary relationships among different hot subdwarf types, and study their binary natures.  

\subsection{Summary}
In this study, we selected 257 composite and 203 single-lined hot subdwarf candidates from LAMOST DR8 dataset with the help of \textit{Gaia} EDR3 H-R diagram and the catalog of \citet{2019A&A...621A..38G}. 
222 stars were confirmed as hot subdwarfs, among which 131 objects were identified with composite spectra, and 91 objects with single-lined spectra. 
We cross-matched our sample with the hot subdwarf star catalog of \citet{2020A&A...635A.193G}, and confirmed that 74 hot subdwarf stars in our sample are new discoveries. 
Though 148 stars have been recorded before, 109 of them were reported without atmospheric parameters.
We obtained their atmospheric parameters by fitting the H and He line profiles with  synthetic spectra. 
Composite hot subdwarfs are well separated from single-lined hot subdwarfs in the two-color diagram. 
Two wide sdB binaries with composite spectra were discovered, sdB\_b1 and sdB\_b2. Their long orbital periods (i.e., $884\pm25$ and $821\pm43$ days, respectively) make them belong to the main group of known wide sdB binaries. However, sdB\_b1 shows the highest eccentricity among these binaries, which is far beyond model predictions. Further study is expected before the nature of period-eccentricity for wide sdB binaries is clearly understood. 
Our sample reported here are  valuable resources to carry out spectroscopic follow-up observations, especially for the 15 composite sdB stars falling in the highly reliable binary region of RUWE-AEN plane. A distinct gap is clearly presented among temperatures of cool companions for the composite sample, but it is a little early to come into the conclusion that this feature is related to the formation history of hot subdwarf stars before their binary natures are confirmed. 

\begin{acknowledgments}
We thank the anonymous referee for his/her valuable comments and useful suggestions, which help improve the manuscript greatly. 
This work acknowledges supports from National Natural Science Foundation of China (Nos. 12073020 and 12273055), 
Scientific Research Fund of Hunan Provincial Education Department 
 grant No. 20K124, Cultivation Project for LAMOST Scientific Payoff and Research Achievement of CAMS-CAS, the science research grants from the China Manned Space Project with No. CMS-CSST-2021-B05. 
 K.Hu acknowledges the support of the Joint Research Funds in Astronomy (U1931115) under cooperative agreement between the National Natural Science Foundation of China and the Chinese Academy of Sciences. 
P.N. and J.V. acknowledge support from the Grant Agency of the Czech Republic (GA\v{C}R 22-34467S) and from the Polish National Science Centre under projects UMO-2017/26/E/ST9/00703 and UMO-2017/25/B/ST9/02218. 
The Astronomical Institute in Ond\v{r}ejov is supported by the project RVO:67985815.
This research has used the services of \mbox{\url{www.Astroserver.org}} under reference OKV6NF, ATGCXY and K4EWOK. Guoshoujing Telescope (the Large Sky Area Multi- Object Fiber Spectroscopic Telescope LAMOST) is a National Major Scientific Project built by the Chinese Academy of Sciences. Funding for the project has been provided by the National Development and Reform Commission. LAMOST is operated and managed by the National Astronomical Observatories, Chinese Academy of Sciences.

\software{astropy \citep{2013A&A...558A..33A,2018AJ....156..123A}, TOPCAT \citep{2005ASPC..347...29T}}

\end{acknowledgments}

%




\bibliography{sd_composite_2022} 
\bibliographystyle{aasjournal}



\end{document}